\documentclass[12pt]{article}

\usepackage{amsmath}
\usepackage{amssymb}
\usepackage{mathrsfs}
\usepackage[utf8]{inputenc}
\usepackage[T1]{fontenc}
\usepackage{amsthm}
\usepackage{enumerate}
\usepackage{fullpage}
\usepackage{multirow}
\usepackage{color}

\usepackage{fullpage}

\newcounter{DDCounter}\addtocounter{DDCounter}{1}

\def\eqn{equation}
\def\cond{condition}
\def\tfn{transformation}

\def\soln{solution}
\def\fn{function}
\def\sm{sigma model}

\def\eqn{equation}

\def\cond{condition}

\def\tfn{transformation}

\def\soln{solution}
\def\fn{function}
\def\sm{sigma model}

\def\dd{Drinfeld double}
\def\mt{Manin triple}

\def\4diml{four-dimensional}
\def\bkg{background}
\def\wrt{with respect to}

\def\-1{^{-1}}
\def\half{\frac{1}{2}}

\def\cd{{\mathfrak d}}
\def\cg{{\mathfrak g}}
\def\tcg{\tilde{\mathfrak g}}

\def\wt{\tilde}
\def\wh{\widehat}

\def\sm{sigma model}
\def\PL{Poisson--Lie }
\def\pltp{Poisson--Lie T-pluralit}

\def\gsugra{Generalized Supergravity Equation}
\def\usugra{standard Supergravity Equation}
\def\sugra{Supergravity Equation}
\def\ybdn{Yang--Baxter deformation}
\def\yb{Yang--Baxter\ }
\def\mr{\mathbf{r}}

\def\cf{{\mathcal {F}}}

\newcommand{\izo}{\longrightarrow}

\newcommand{\unit}{\mathbf{1}}
\newcommand{\nul}{\mathbf{0}}

\newcommand{\D}{\mathscr{D}}
\newcommand{\G}{\mathscr{G}}
\newcommand{\tG}{\widetilde{\mathscr{G}}}

\newcommand{\cB}{\mathcal B}
\newcommand{\cD}{\mathcal D}
\newcommand{\cF}{\mathcal F}
\newcommand{\cG}{\mathcal G}
\newcommand{\cJ}{\mathcal J}

\newcommand{\cR}{\mathcal R}

\newcommand{\cW}{\mathcal W}

\begin{document}
\title{Eight-dimensional Manin triples, Yang--Baxter deformations and solutions of Supergravity Equations}
\author{Ladislav Hlavat\'y\footnote{ladislav.hlavaty@fjfi.cvut.cz}, Petr Novotn\'y\footnote{petr.novotny@fjfi.cvut.cz}
%\\ {\em Department of Physics,}
\\ {\em Faculty of Nuclear Sciences and Physical Engineering,}
\\ {\em Czech Technical University in Prague,}
\\ {\em B\v rehov\' a 7, 115 19, Prague 1,}
\\ {\em Czech Republic}
\and
Ivo Petr\footnote{ivo.petr@fit.cvut.cz}
%\\ {\em Department of Applied Mathematics,}
\\ {\em Faculty of Information Technology,}
\\ {\em Czech Technical University in Prague,}
\\ {\em Th\' akurova 9, 160 00, Prague 6,}
\\ {\em Czech Republic}
}
\maketitle
\begin{abstract}
Extensive list of 4+4-dimensional Manin triples that was presented recently can be used to find new solutions of supergravity equations via Poisson--Lie T-plurality. To get the solutions we start with 1+3-dimensional flat backgrounds on Poisson--Lie groups corresponding to semi-Abelian Manin triples. For application of the Poisson--Lie T-plurality we identify Manin triples that form various decompositions of the same Drinfeld double. Beside flat backgrounds and plane-parallel waves solving supergravity equations, plurality transformation also produces curved backgrounds with torsion satisfying (generalized) supergravity equations. Many of the Poisson--Lie transformations can be understood as homogeneous Yang--Baxter deformations. Of special interest are the non-unimodular deformations leading to solutions of generalized supergravity equations.
\end{abstract}
 
%\keywords{Sigma Models; \dd s; Poisson-Lie T-duality/plurality, Supergravity Equations, Yang-Baxter deformations.}

%\ccode{PACS numbers: 11.10.Lm, 11.27.+d,  04.60.Cf, 02.30.Ik}

\tableofcontents

%%%%%%%%%%%%%%%%%%%%%%%%%%%%%%%%%%%%%%%%%%%%%%%%%%%%%%%%%%%%%%%%%%%%%%%%%%%%%%%
%% Introduction
%%%%%%%%%%%%%%%%%%%%%%%%%%%%%%%%%%%%%%%%%%%%%%%%%%%%%%%%%%%%%%%%%%%%%%%%%%%%%%%
\section{Introduction}
Consistency of string theory at quantum level requires Weyl invariance. In the lowest order of $\alpha'$-expansion this means that the \bkg\ fields have to satisfy vanishing beta function equations \cite{fratsey, calmar} called also \sugra s. Solving these equations is rather complicated in general. One of the tools that can be used to generate new solutions is \PL T-duality or plurality \cite {klise,klim:proc,unge} that was discussed and employed as a solution-generating technique e.g. in \cite{sibylla, saka2, hlape:bianchisugra}.

Research in non-Abelian T-duality has shown that dualization with respect to non-semisimple Lie groups does not preserve Weyl invariance. When the trace $f_b{}^{ba}$ of the structure constants of the corresponding Lie algebra does not vanish, the resulting \bkg s possess a mixed gauge and gravitational anomaly, and the resulting backgrounds do not always satisfy the \usugra s \cite{grv, aagl}. Similar problem appeared in the study of integrable deformations until \sugra s were modified to \gsugra s \cite{sugra2, Wulff:2016tju}. 

As certain deformations can be understood within the framework of \PL T-duality \cite{hoaretsey, ostong, BW1806.04083} and Double Field Theory \cite{Sakamoto:2017cpu, LuOst, dehato}, it turns out that T-dual/plural models satisfy these modified equations as well. In the notation of \cite{hkc} the \gsugra s for bosonic fields in the NS-NS sector read
\begin{align}\label{betaG}
0 &= R_{\mu\nu}-\frac{1}{4}H_{\mu\rho\sigma}H_{\nu}^{\
\rho\sigma}+\nabla_{\mu}X_{\nu}+\nabla_{\nu}X_{\mu},\\ \label{betaB}
0 &=
-\frac{1}{2}\nabla^{\rho}H_{\rho\mu\nu}+X^{\rho}H_{\rho\mu\nu}+\partial_{\mu}X_{\nu}-\partial_{\nu}X_{\mu},\\
\label{betaPhi} 0 &=
R-\frac{1}{12}H_{\rho\sigma\tau}H^{\rho\sigma\tau}+4\nabla_{\mu}X^{\mu}-4X_{\mu}X^{\mu}.
\end{align}
Here $R_{\mu\nu}$ and $R$ are Ricci tensor and scalar curvature of metric $\cG$ specifying sigma-model \bkg\ $\cf= \cG+\cB$ together with Kalb-Ramond field $\cB$, which gives rise to torsion $H = \mathrm{d}\cB$.
Components of the one-form $X$ are calculated from dilaton $\Phi$ and Killing vector field $\cJ$ as
\begin{equation*}\label{xform}
X_\mu:=\partial_\mu\Phi+\cJ^\kappa\cf_{\kappa\mu}.
\end{equation*}
The condition that $\cJ$ is a Killing vector field of the \bkg\ $\cf$ is not necessary for our investigation of eqns. \eqref{betaG}--\eqref{betaPhi}, but it is required for their full version containing the R-R fields.

For $\cJ=0$ the \gsugra s become \usugra s. However, $X$ is invariant \wrt\ gauge \tfn\
\begin{equation}\label{gauge tfn lambda}
\Phi_\Lambda:=\Phi +\Lambda, \quad \cJ_\Lambda:= \cJ-\mathrm{d}\Lambda \cdot \cf\-1
\end{equation}
for arbitrary differentiable function $\Lambda$. As a result, if $\mathrm{d}X=0$, we can always find dilaton $\Phi_\Lambda$ such that $X_\mu=\partial_\mu\Phi_\Lambda$ and  $\cJ_\Lambda=0$. The \eqn s \eqref{betaG}--\eqref{betaPhi} then again reduce to \usugra s. On the other hand, it follows from the \eqn\ \eqref{betaB} that $\mathrm{d}X=0$ whenever the torsion $H$ vanishes. If $\mathrm{d}X\neq 0$, we cannot trivialize the Killing vector field $\cJ$ by gauge \tfn\  and the \eqn s \eqref{betaG}--\eqref{betaPhi} are true \gsugra s. Examples of solutions with non-trivial $\cJ$ were obtained e.g. in Refs. \cite{saka2,hlape:bianchisugra} via \PL T-plurality with respect to three-dimensional groups.

In an earlier work \cite{hlapet:cqgrav} we investigated non-Abelian T-duality of the flat Minkowski \bkg. Duality was carried out with respect to four-dimensional subgroups of symmetries generated by subalgebras\footnote{Three-dimensional algebras are denoted according to Bianchi classification. For four-dimensional indecomposable algebras we use the notation of \cite{snowin:kniha}.}
\begin{equation}\label{alryPWZ}
B3+A_1, B4+A_1, B5+A_1, B6_0+A_1, B7_0+A_1, s_{4,11}, s_{4,3}^{1,1}, s_{4,5}^{a,a}, s_{4,5}^{a,0}
\end{equation}
of the Poincaré algebra classified\footnote{Even though it seems that there is much more four-dimensional subalgebras in Ref. \cite{PWZ} than here,  many of them are just various realizations of \eqref{alryPWZ} in terms of generators of the Poincaré algebra. Corresponding \sm s then differ by the choice of the constant matrix $E_0$ introduced in \eqref{metr}. Moreover, not all of the subalgebras in the Ref. \cite{PWZ} produce solutions with nontrivial scalar curvatures that we are looking for.} in \cite{PWZ}. Together with four-dimensional Abelian algebra $A_4$ these algebras form semi-Abelian \mt s $( \mathfrak g | A_4 ) $ that give us the underlying algebraic structure of \PL T-duality and plurality. The classification of 4+4-dimensional "standard" Manin triples $(\mathfrak g|\tcg)$ presented in \cite{HNP} allows us to investigate not only duality but also full \PL T-plurality. We can then generate much richer set of \bkg s. Most of them turn out to be flat or plane-parallel\footnote{pp-wave metrics can be recognized by vanishing curvature scalars and verifying that there exists a covariantly constant null Killing vector of the metric.4}wave metrics with or without torsion that solve the \usugra s.  Nevertheless, there are also \bkg s with non-vanishing scalar curvature and torsion. We focus on them in the following as they provide a chance to get solution of the \gsugra s.

Integrable sigma models and their \yb deformations \cite{klimyb2, BW1609} were extensively studied in the past. Since they allow further investigation of the AdS/CFT correspondence, the focus has been mainly on the deformations of $AdS_n \times S^n$-type \bkg s \cite{Araujo:2017jkb, Araujo:2017jap, BW1806.04083}. It is, however, interesting to study also deformations of other \bkg s, such as the flat \bkg\ \cite{IdiabTong, Idiab} or WZW sigma model \cite{EgPar1,EgPar2}. We shall show that many of the 4+4-dimensional \mt s in \cite{HNP} produce \yb deformations of the flat Minkowski metric. These deformations follow from both unimodular and non-unimodular $R$-matrices so that one can observe the influence of unimodularity on the type of \sugra s that the deformed \bkg s satisfy, cf. \cite{IdiabTong2, Hronek:2020skb}.

The structure of the paper is as follows. In Sections \ref{examples preDD} and \ref{sec:PLTP} we review the notion of \dd, \mt, and summarize \PL T-plurality formulas. In Section \ref{sec:ex_nonzero_R} we focus on \ybdn s that lead to \bkg s with nonzero scalar curvature. In the non-unimodular cases we obtain solutions of \gsugra s. In Section \ref{nonunimpp} we give further examples of non-unimodular \ybdn s leading to plane-parallel waves.

\section{Drinfeld double equivalence of Manin triples}\label{examples preDD}

\dd\ $\D=(\G|\tG)$ is a real $2D$-dimensional Lie group whose Lie algebra $\cd$ decomposes into direct sum of two $D$-dimensional subalgebras $\cg$ and $\tcg$. Moreover, the Lie algebra $\cd$ is equipped with an ad-invariant non-degenerate symmetric bilinear form $\langle . , . \rangle$ and the subalgebras $\cg$ and $\tcg$ are maximally isotropic with respect to $\langle . , . \rangle$.

In the subalgebras $\cg$ and $\tcg$ we may choose bases $T_a \in \cg,\ \widetilde{T}^a \in \tcg$, $a=1, \ldots, D,$ satisfying
\begin{align*}
\label{dual_bases}
\langle T_a, T_b \rangle & = 0, & \langle \widetilde{T}^a, \widetilde{T}^b \rangle &= 0, & \langle T_a, \widetilde{T}^b \rangle & = \delta_a^b.
\end{align*}
The algebraic structure of the Manin triple is then given by commutation relations of the subalgebras $\cg$ and $\tcg$,
\begin{equation*}
\label{commutation_on_d}
[T_a,T_b]=f_{ab}{}^c T_c, \qquad [\widetilde T^a, \widetilde T^b]=\wt f^{ab}{}_c \widetilde T^c, \qquad [T_a,\widetilde T^b]=f_{ca}{}^b \widetilde T^c + \wt f^{bc}{}_a T_c.
\end{equation*}

\pltp y is based on the \dd\ isomorphisms between Manin triples. Two Manin triples $MT,MT'$  with bases $T_A = (T_a,\widetilde{T}^a)$, $T'_A = (T'_a,\widetilde{T}'^a)$ and defining relations 
$$
[T_A,T_B]=X_{AB}{}^C T_C,\quad [T'_{A},T'_{B}]=X'_{AB}{}^C T'_{C}\quad A,B,C = 1,\ldots,2D
$$
are \dd\ isomorphic if and only if there is an invertible $2D\times 2D$ matrix $M$ relating the bases as
$$ T'_A=M_A{}^B\, T_B,$$
which appropriately transforms their algebraic structure, i.e. satisfies
\begin{equation}
X'_{AB}{}^C=M_A{}^D M_B{}^E X_{DE}{}^F(M\-1)_F{}^C,\label{tfnX}
\end{equation}
and preserves the bilinear form $\langle . , . \rangle$, i.e.
\begin{equation}
M_A{}^{D}M_B{}^{E}\eta_{DE}= \eta_{AB},\qquad \eta_{AB}=\left(\begin{array}{cc} \mathbf{0}_D & \unit_D\\  \unit_D & \mathbf{0}_D \\ \end{array}\right).\label{ODD}
\end{equation}

Complete classification of Manin triples composed of four-dimensional algebras proved to be difficult. Therefore, in \cite{HNP} we restricted our considerations to \mt s in some "standard" form. Even with this restriction we obtained extensive list of nearly 200 \mt s. Classification of corresponding \dd s is even harder. Nevertheless, for \mt s containing subalgebras of the Poincaré algebra  \eqref{alryPWZ} we were able to sort \mt s into classes according to their Lie algebra invariants, and find the corresponding matrices $M$, thus proving that the \mt s are \dd\ isomorphic.

We should note that matrices $M$ that solve the \eqn s \eqref{tfnX} and \eqref{ODD} are determined up to automorphisms of the algebras that constitute the \mt s.  This arbitrariness can be used to choose $M$ in a simple form. Sets of isomorphic \mt s used in the paper together with the corresponding matrices $M$ are given in the Appendix.

As mentioned in Ref. \cite{kurmus}, many of the 4+4-dimensional \mt s can be obtained from a semi-Abelian \mt \ $(\mathfrak g\,| A_4)$ by \tfn\ of bases
\begin{equation}\label{ybtfn}
T'_a=T_a, \qquad \widetilde T'^a= r^{ab} T_b+\widetilde T^a,
\end{equation} 
where $r^{ab}$ are components of antisymmetric matrix $\mathbf r$ satisfying homogeneous classical \yb \eqn\ in $\cg$
\begin{equation*}
\label{cybe}
[\mr_{12},\mr_{13}]+[\mr_{12},\mr_{23}]+[\mr_{13},\mr_{23}]=0.
\end{equation*}
Using the structure constants $f_{km}{}^n$ we can write it as
\begin{equation}
\label{cybe2}
\left(\,r^{ik}r^{jm}f_{km}{}^n+r^{ik}r^{nm}f_{mk}{}^j+r^{jk}r^{nm}f_{km}{}^i\,\right)\, T_i\otimes T_j\otimes T_n=0.
\end{equation}
In the following we shall see that many of the standard \mt s are indeed related by matrices
\begin{equation} \label{mpqrs}
M=\left(\begin{array}{cc} \unit_4 & \mathbf{0}_4\\  R & \unit_4 \\ \end{array}\right).
\end{equation}
where $R$ satisfies the classical \yb \eqn. \pltp ies that follow from these \dd\ isomorphisms then produce \yb deformed \bkg s.

\section{\pltp y formulas}\label{sec:PLTP}

The \bkg\ of a \PL \sm\ is given by covariant tensor field $\cf$ whose symmetric and antisymmetric part represent metric and Kalb-Rammond field on a Lie group $\G$. It can be called \PL symmetric if the Lie derivatives of $\cf=\cG+\cB$ with respect to left-invariant vector fields ${V_a}$ of the group $\G$ satisfy the \cond\ \cite{klise}
\begin{equation}\label{KScond}
({\cal L}_{V_a}\cf)_{\mu \nu}~=~
 -\cf_{\mu \rho} ~{V_b}^{\rho}\;\tilde
f^{cb}{}_{a}\;{V_c}^{\lambda}\;\cf_{\lambda \nu}
\end{equation}
for structure constants ${\tilde f}^{cb}{}_{a}$ of some Lie algebra ${\mathfrak{\tilde g}}$. The self-consistency of the condition \eqref{KScond} implies that algebras $\cg$ and $\tcg$ form a \mt.

The general solution of the equation \eqref{KScond} that we shall use in the following is
\begin{equation}\label{met}
\cf_{\mu\nu}(x)={r_{\mu}}^{a}(g(x))\, E_{ab}(g(x)) \, {r_{\nu}}^{b}(g(x)),
\end{equation}
where ${r_{\mu}}^{a}(g(x))$ are the components of right-invariant forms $dg g^{-1}$ expressed in coordinates $x^\mu$ on the group $\G$,
\begin{equation}\label{metr}
E(g(x))=\left(E_{0}^{-1}-\Pi(g(x))\,\right)^{-1}, \qquad \Pi(g)=b(g) \cdot a(g)^{-1},
\end{equation}
$E_0$ is constant invertible matrix, and matrices $a(g),b(g)$ are given by the adjoint representation of the Lie group $\G$ on the Lie algebra of the Drinfeld double $\cd$. For details see e.g. \cite{klim:proc}.

If we want to satisfy the \sugra s \eqref{betaG}--\eqref{betaPhi} for given \bkg\ $\cf$ we must find  dilaton $\Phi$ and Killing vector $\cJ$. Denoting the components of the left-invariant form $g\-1dg$ as $l_{\mu}{}^a$ the dilaton $\Phi$ can be also written as
\begin{align} \Phi =\varphi +\frac{1}{4}\,\ln\left(\frac{\det \cG}{\det(l_{\mu}{}^a)^2}\right),
\label{eq:dilaton-JL}
\end{align}
and knowledge of \fn\ $\varphi$ is complementary to knowledge of the dilaton. Nevertheless, when we start with flat initial model where $\cJ=0$ and $\Phi = 0$, we can use this formula for getting the initial $\varphi$.

\pltp y transforms the initial model to a new one. The transformed  $E_0$ reads 
\begin{equation}\label{E0hat}
\wh E_0=(Q+ P \cdot E_0 ) \cdot
(S+R\cdot E_0)\-1,
\end{equation}
where $P,Q,R,S$ are the $D\times D$ blocks of the matrix \eqref{mpqrs} that satisfies \eqref{tfnX}, \eqref{ODD}, and relates the \mt\ $MT=(\cg|\tcg)$ to \dd\ isomorphic \mt\ $MT'=(\hat \cg|\bar\cg)$. The \PL plural \bkg\ on the group $\wh{\mathscr{G}}$ is then obtained analogously to \eqref{met} and \eqref{metr}. For the \ybdn\ we introduce deformation parameter $\eta$ by $R \rightarrow \eta R$ and the transformed $\wh E_0$ simplifies to
\begin{equation}\label{ybE0}
\wh E_0=(E_0\-1+\eta\,R)^{-1}.
\end{equation}

Transformation of the dilaton is rather intricate. We shall adopt the approach of Ref. \cite{saka2} which uses the formalism of Double Field Theory. For the DFT dilaton 
$$d=\Phi-\frac{1}{2}\ln(\sqrt{|\det\cG|})$$
and generalized vielbein written in terms of the right-invariant vector fields $e_a$ and one-forms $r^a$ as
\begin{equation*}
E_A{}^M = \begin{pmatrix} e_a{}^m & 0 \\ \Pi^{ab}\,e_b{}^m & r^a{}_m \end{pmatrix}
\end{equation*}
we define the flux
\begin{align}\label{FM}
\cF_A = \cW^B{}_{AB} + 2\, \cD_A d  
\end{align}
through\footnote{DFT indices are raised and lowered by $\eta^{AB} = \eta_{AB}$.}
\begin{equation*}
\cW_{ABC} \equiv - \cD_A E_B{}^M\, E_{MC}, \qquad 
\cD_A \equiv E_A{}^M \partial_M.
\end{equation*}
For the initial model, where $\tilde f_b{}^{ba}=0$, the corresponding $\cF_M$ associated with $\cF_A$ via $\cF_A= E_A{}^M\, \cF_M$ can be calculated as
\begin{align*}\label{gradfi}
\cF_M= 2\, \partial_M \varphi .
\end{align*}
To ensure that the fluxes $\cF_A$ transform covariantly under the \dd\ isomorphism, i.e.
\begin{equation}\label{hatFA}
\wh \cF_A=M_A{}^B \cF_B,
\end{equation}
we have to perform the shift \cite{dehath,saka2}
$$\partial_M d\mapsto  \partial_M d+\begin{pmatrix} \nul \\ \cJ^m\end{pmatrix}.$$
The \PL transformed dilaton is found from function $\wh \varphi$ that, together with the Killing field $\wh\cJ$, follows from the transformed flux $\wh \cF_M=(\wh E\-1)_M{}^A\wh \cF_A $ as 
\begin{align}\label{rhs}
 \begin{pmatrix} \partial_m \wh \varphi(x) \\  \wh \cJ^m(x)
\end{pmatrix}
=&\half\wh \cF_M  + \begin{pmatrix} \nul \\ \half \bar f_b{}^{ba} \,\hat V_a^m\end{pmatrix},
\end{align}
where $ \hat V_a^m$ are components of left-invariant vector fields of the group associated with $\hat{\cg}$.

\section{Solutions of \sugra s with non-vanishing scalar curvature}\label{sec:ex_nonzero_R}

For given $\cf$ one can try to solve equations \eqref{betaB}--\eqref{betaPhi} for the one-form $X$. Rather helpful are the compatibility equations for derivatives
\begin{align}\label{dmuX}
\partial_\mu X_{\nu}&=\half(X_\rho A^\rho{}_{\mu\nu}+W_{\mu\nu}), 
\end{align}
that follow from \eqref{betaG} and \eqref{betaB}, where
$$ A^\rho{}_{\mu\nu} = \Gamma^\rho{}_{\mu\nu}-H^\rho{}_{\mu\nu},  \quad W_{\mu\nu}=
\half\nabla_\rho H^\rho{}_{\mu\nu}+\frac{1}{4}H_{\mu\sigma\rho}H_\nu{}^{\sigma\rho}-R_{\mu\nu}.$$
They yield linear inhomogeneous equations for $ X_{\mu}$
\begin{equation}\label{linearX}
X_\kappa\, C^\kappa{}_{[\lambda\mu]\nu}+D_{[\lambda\mu]\nu}=0,
\end{equation}
where
$$
 C^\kappa{}_{\lambda\mu\nu}=A^\kappa{}_{\lambda\rho}A^\rho{}_{\mu\nu}+\partial_\lambda A^\kappa{}_{\mu\nu}, \quad D_{\lambda\mu\nu}=\partial_\lambda W_{\mu\nu}+W_{\lambda\rho}A^\rho{}_{\mu\nu}.
$$
Solution of linear \eqn s \eqref{linearX} reduces the number of unknown components of the one-form $X$ and facilitates the  solution of \eqref{dmuX}. 

Another tool for solving \sugra s is the \pltp y. Looking for nontrivial \soln s we can start with a simple solution, e.g. (Ricci) flat metric without torsion, and apply the \PL \tfn\ to obtain new \bkg\ that solves (generalized) \sugra s. To do so, we must first express the flat metric as the metric on a four dimensional \PL group that represents its group of symmetries. That means finding the adapted coordinates, as we did in \cite{hlapet:cqgrav}, or finding constant $E_0$ such that $\cF$ gives (Ricci) flat \bkg. In this paper we choose the latter approach. The plural models then follow from the \dd\ isomorphisms of the semi-Abelian \mt\ $(\mathfrak g|A_4)$ where $\mathfrak g$ is the Lie algebra of the group of symmetries of the flat metric.

\subsection{\yb deformations}

From the list of \dd s in the Appendix we can see that many of the \mt s $MT'$ are \dd\ isomorphic to $MT=(\mathfrak g| A_4)$ via \eqref{mpqrs}. Therefore, after introducing parameter $\eta$ we can use the modified matrix
\begin{equation*}\label{mR}
 M=\left(\begin{array}{cc} \unit_4 & \bf{0}_4 \\  
 \eta\,R & \unit_4 \\ \end{array}\right)
\end{equation*}
to get the $\eta $-deformed \yb models by the \pltp y. Alternatively, the deformed \bkg s can be calculated by a simpler formula \cite{BW1806.04083}
\begin{equation*}
\wh\cG - \wh\cB =(\cG - \cB)\left(\unit-\eta\,\Theta\cdot(\wh\cG - \wh\cB)\right)\-1,
\end{equation*}
where
\begin{equation*}
\Theta^{\mu\nu}= V_a{}^\mu \, R^{ab} \,  V_b{}^\nu.
\end{equation*}
and $V_a$ 
are the left-invariant vector fields. In the literature $\Theta$ is referred to as non-commutative structure \cite{Araujo:2017jkb, Araujo:2017jap} and is used to calculate components of vector $\cJ$ as 
\begin{equation}\label{def_J}
\nabla_{\mu}\Theta^{\mu\nu}=\cJ^{\nu}.
\end{equation}
We have verified that these formulas hold for all the \ybdn s considered. Nevertheless, since \eqref{def_J} does not give correct $\cJ$ for general \pltp y \cite{hlape:bianchisugra}, we use the transformation rules given in Section \ref{sec:PLTP} for our calculations.

The transformation \eqref{hatFA} leads to a shift of the flux $\mathcal F_A$
$$
\wh{\cF_A} = M_A{}^B \cF_B = \cF_A-K_A,
$$ 
where $K_A = E_A{}^M K_M$ and $K_M=(0,K^m)$, see \cite{BLW2003.05876,Hronek:2020skb}, with
\begin{equation*}
K^m=-\eta\,\Omega^c\, V_c^m
%\label{verKm}
\end{equation*}
and
$$\Omega^c=\,R^{ab}f_{ab}{}^c. $$
The issue of the so called unimodularity of the matrix $R$ rises here. If the $R$-matrix is unimodular, i.e. if $\Omega^c=0$, then the deformation gives solution of \usugra s. If not, we can get solutions of either standard or  Generalized \sugra s.

Maybe surprisingly, for the \yb deformations of flat models the vector $K^m$ is equal to the Killing vector
\begin{equation*}
\label{Killing} \wh\cJ^m=\half (\wh {\mathcal F}^m+\bar f_b{}^{ba}V_a{}^m).
\end{equation*}
This is equivalent to the identity 
\begin{equation}
\label{identity}\left(\frac{1}{\eta}b(g)^{ab}-a(g)^{b}{}_cR^{ca}\right)\mathcal F_a - R^{ba}f_{ad}{}^d=0
\end{equation} 
that holds in all examples of \yb deformations we have investigated.

Indeed, the structure constants of the \yb deformed \mt\ are 
\begin{equation*}
\label{barf} \bar f_a{}^{bc}=-\eta\left(R^{bd}f_{ad}{}^c - R^{cd}f_{ad}{}^b \right),
\end{equation*}
so that
\begin{equation*}
\label{trbarf} \bar f_b{}^{ba}=-\eta\left(R^{cd}f_{cd}{}^a - R^{ac}f_{bc}{}^b\right)= \eta\left(\Omega^a+ R^{ac}f_{cb}{}^b\right).
\end{equation*}
Moreover,
\begin{equation*}
\label{calFM} \wh {\cF}_M =U_M{}^A \wh{\cF}_A = U_M{}^A( {\cF}_A-K_A)
\end{equation*}
where the matrix $U_M{}^A$ is the inverse of $E_A{}^M$ equal to
\begin{equation*}
 U_M{}^A = \begin{pmatrix} r_m{}^a & 0 \\ -e^m{}_b\Pi^{ba} &e^m{}_a \end{pmatrix}.
\end{equation*}
Therefore,
\begin{equation}
\label{kseqkill} \cJ^m-K^m=\frac{\eta}{2}\left[ e^m_c(-\pi^{ca}-R^{ca})\mathcal F_a -V^m{}_b\, R^{ba}\, f_{bc}^c\right],
\end{equation}
and using $e_c{}^mV_m{}{}^b= a_c{}^b$ we find that the r.h.s. of \eqref{kseqkill} is proportional to the l.h.s. of \eqref{identity}.

\subsection{Unimodular \yb deformations}\label{YangBaxter}

\subsubsection{Pluralities of {$(s_{2,1}+A_2|A_4)$}}

As the first example let us consider semi-Abelian Manin triple denoted $(s_{2,1}+A_2|A_4)$, and the sigma model on the Lie group whose Lie algebra $\cg=s_{2,1}+A_2 \cong B3+A_1$ has commutation relations
$$[T_1,T_2]=T_2.$$ 
In this and all the following examples we shall parametrize group elements $g(x)$ as
$$
g(x)=e^{x_1 T_1}e^{x_2 T_2}e^{x_2 T_2}e^{x_3 T_3}e^{x_4 T_4},
$$
where we lowered the coordinate indices for the sake of clarity of the expressions appearing in the \bkg\ fields.

It is rather straightforward to use formulas \eqref{met}--\eqref{eq:dilaton-JL} to construct the background fields for given $E_0$ and $\varphi$. Since we are interested in backgrounds satisfying \eqref{betaG}--\eqref{betaPhi}, we must first find appropriate $E_0$ and $\varphi$. A possible approach is to find these data first for (Ricci) flat background with vanishing torsion and dilaton, and then use the formulas \eqref{E0hat} and \eqref{rhs} to get $\wh E_0, \wh \varphi$ and $\wh \cJ$ for more complicated backgrounds satisfying the (generalized) \sugra s. 

For a semi-Abelian Manin triple the Poisson bivector $\Pi$  in \eqref{metr} vanishes. We may then choose $E_0$ symmetric to obtain \bkg\ with vanishing $\cB$-filed and torsion $H$. With this ansatz it is possible to find appropriate $E_0$ and construct flat model for the Manin triple $(s_{2,1}+A_2|A_4)$. Its general form is
$$\cF(x)=\cG(x) = \left(
\begin{array}{cccc}
 E_{11} & E_{12}\, e^{x_1} & E_{13} & E_{14} \\
 E_{12}\, e^{x_1} & 0 & 0 & 0 \\
 E_{13} & 0 & E_{33} & E_{34} \\
 E_{14} & 0 & E_{34} & E_{44} \\
\end{array}
\right).$$
Moreover, we can choose $E_{12}=E_{33}=E_{44}=1$ and other $E_{ij}=0.$ To get vanishing dilaton $\Phi=0$ from \eqref{eq:dilaton-JL} we choose $\varphi = -\frac{x_1}{2}$. 

This way we have constructed a solution of \sugra s corresponding to the the semi-Abelian Manin triple $(s_{2,1}+A_2|A_4)$ that is \dd\ isomorphic to other Manin-triples, namely
\begin{align*}
& (s_{2,1}+A_2|A_4)  \cong (s_{2,1}+A_2|B2+A_1; \operatorname{P1})\cong\\ & \cong (s_{2,1}+A_2|s_{2,1}+A_2; \operatorname{P7}) \cong (s_{2,1}+A_2|s_{2,1}+A_2; \operatorname{P15}).
\end{align*}
These represent various decompositions of the  Drinfeld double into standard Manin triples. The Lie algebra invariants of these Manin triples and \tfn s among them are presented in the Appendix. For this \dd\ all the isomorphisms $M$ solving \eqref{tfnX}, \eqref{ODD} can be chosen to have a simple form of a \ybdn.

Except for one case we find that the plural models are flat and torsionless. The only model with non-vanishing scalar curvature is obtained by the \tfn\  $(s_{2,1}+A_2|A_4)\izo (s_{2,1}+A_2|s_{2,1}+A_2;\operatorname{P15})$. Introducing the deformation parameter $\eta$ it is represented by the matrix
$$
M=\left(
\begin{array}{cccccccc}
 1 & 0 & 0 & 0 & 0 & 0 & 0 & 0 \\
 0 & 1 & 0 & 0 & 0 & 0 & 0 & 0 \\
 0 & 0 & 1 & 0 & 0 & 0 & 0 & 0 \\
 0 & 0 & 0 & 1 & 0 & 0 & 0 & 0 \\
 0 & 0 & \eta & 0 & 1 & 0 & 0 & 0 \\
 0 & 0 & 0 & 0 & 0 & 1 & 0 & 0 \\
-\eta & 0 & 0 & 0 & 0 & 0 & 1 & 0 \\
 0 & 0 & 0 & 0 & 0 & 0 & 0 & 1 \\
\end{array}
\right).
$$
The $\eta$-paramatrized form of the Manin triple $(s_{2,1}+A_2|s_{2,1}+A_2,\operatorname{P15})$ is given by commutation relations of four-dimensional algebras $\mathfrak{g}$ and $\tilde{\mathfrak{g}}$ as
$$
[T_1,T_2]=T_2, \quad [\widetilde{T}^2, \widetilde{T}^3]= -\eta\,\widetilde{T}^2.
$$
This is a \ybdn\ of the Manin triple $(s_{2,1}+A_2|A_4)$ given by unimodular $R$ matrix. Using formulas \eqref{met}--\eqref{ybE0} we get plural \bkg
$$\cf(x)= \cG(x)+\cB(x) = \left(
\begin{array}{cccc}
 \frac{\eta ^2 e^{2 x_1} x_2^2}{2 \eta ^2 e^{x_1} x_2-1} & \frac{e^{x_1}
   \left(\eta ^2 e^{x_1} x_2-1\right)}{2 \eta ^2 e^{x_1} x_2-1} & \frac{\eta 
   e^{x_1} x_2}{1-2 \eta ^2 e^{x_1} x_2} & 0 \\
 \frac{e^{x_1} \left(\eta ^2 e^{x_1} x_2-1\right)}{2 \eta ^2 e^{x_1} x_2-1} &
   \frac{\eta ^2 e^{2 x_1}}{2 \eta ^2 e^{x_1} x_2-1} & \frac{\eta  e^{x_1}}{2
   \eta ^2 e^{x_1} x_2-1} & 0 \\
 \frac{\eta  e^{x_1} x_2}{2 \eta ^2 e^{x_1} x_2-1} & \frac{\eta  e^{x_1}}{1-2
   \eta ^2 e^{x_1} x_2} & \frac{1}{1-2 \eta ^2 e^{x_1} x_2} & 0 \\
 0 & 0 & 0 & 1 \\
\end{array}
\right).$$
The metric $\cG$ has scalar curvature
$$
\cR = \frac{2 \eta ^2 \left(4 \eta ^2 e^{x_1} x_2+5\right)}{\left(1-2 \eta ^2 e^{x_1}x_2\right)^2},
$$
and the $\cB$-field has non-trivial torsion $H = d\cB$. Formula \eqref{rhs} gives us vanishing vector field $\cJ$ and dilaton 
$$
\Phi(x)=-\frac{1}{4} \ln \left(1-2 \eta ^2 e^{x_1} x_2\right)^2.
$$
The \bkg\ satisfies the \usugra s and the metric can be diagonalized to the form
$$
ds^2=\frac{1}{1-2\eta\,(t+y_2)}(-dt^2+dy_1^2)+dy_2^2+dy_3^2
$$
by coordinate \tfn\
$$
x_1 = \ln\left(\frac{t+y_2}{\eta}\right)-\eta\,y_2, \quad x_2=e^{\eta y_2}, \quad x_3 = y_1, \quad x_4 = y_3.
$$  

\subsubsection{Pluralities of {$(B4+A_1|A_4)$}}

Commutation relations of the semi-Abelian \dd\ $(B4+A_1|A_4)$ read
$$
[T_1, T_2]=- T_2+ T_3, \quad [T_1, T_3]=- T_3,
$$
and flat models are given by tensor
\begin{equation}\label{mtziniB4}
\cF(x) = \cG(x)=\left(
\begin{array}{cccc}
 E_{11} & e^{-x_1} (E_{12}+E_{13} x_1) & E_{13} e^{-x_1} & E_{14} \\
 e^{-x_1} (E_{12}+E_{13} x_1) & E_{22} e^{-2 x_1} & 0 & 0 \\
 E_{13} e^{-x_1} & 0 & 0 & 0 \\
 E_{14} & 0 & 0 & E_{44} \\
\end{array}
\right). 
\end{equation}
For simplification we can choose $E_{13} = E_{22} = E_{44} = 1$ and the other constants $E_{ij} = 0$. For $\varphi = x_1$ the dilaton vanishes and \sugra s are satisfied.

Standard Manin triples appearing in the \dd\ 3 together with the \mt\ $(B4+A_1|A_4)$ can be found in the Appendix. All but one of the \PL models plural to \eqref{mtziniB4} are pp-waves or flat Minkowski \bkg. The only model with non-vanishing scalar curvature is obtained from the transformation $(B4+A_1|A_4) \izo (B4+A_1|B4+A_1; \operatorname{P24})$. The \ybdn
\begin{equation}\label{izoB4A4B4B4}
M = \left(
\begin{array}{cccccccc}
 1 & 0 & 0 & 0 & 0 & 0 & 0 & 0 \\
 0 & 1 & 0 & 0 & 0 & 0 & 0 & 0 \\
 0 & 0 & 1 & 0 & 0 & 0 & 0 & 0 \\
 0 & 0 & 0 & 1 & 0 & 0 & 0 & 0 \\
 0 & 0 & 0 & \eta  & 1 & 0 & 0 & 0 \\
 0 & 0 & 0 & 0 & 0 & 1 & 0 & 0 \\
 0 & 0 & 0 & 0 & 0 & 0 & 1 & 0 \\
 -\eta  & 0 & 0 & 0 & 0 & 0 & 0 & 1 \\
\end{array}
\right)
\end{equation}
is induced by unimodular R matrix, and the $\eta$-parametrized \mt\ has commutation relations
\begin{align*}
[T_1, T_2] & =- T_2+ T_3, \quad [T_1, T_3] = - T_3,\quad [\widetilde{T}^2, \widetilde{T}^4] = \eta\, \widetilde{T}^2, \quad [\widetilde{T}^3, \widetilde{T}^4] = \eta\,(- \widetilde{T}^2+ \widetilde{T}^3).
\end{align*}
The plural metric and $\cB$-field are rather extensive to display, but can be obtained as inverse of matrix
$$ \cf\-1 =\left(
\begin{array}{cccc}
 0 & 0 & e^{x_1} & \eta  \\
 0 & e^{2 x_1} & -e^{2 x_1} x_1 & \eta  x_2 \\
 e^{x_1} & -e^{2 x_1} x_1 & e^{2 x_1} x_1^2 & \eta  (x_3-x_2) \\
 -\eta  & -\eta  x_2 & \eta  (x_2-x_3) & 1 \\
\end{array}
\right).
$$
The scalar curvature reads
$$
\cR=\frac{2 \eta ^2 e^{2 x_1} \left(-2 \eta ^2 x_2^2-4 \eta ^2 e^{x_1} ((x_1-1)
   x_2+x_3)+\left(7 \eta ^2+5\right) e^{2 x_1}\right)}{\left(\eta ^2 x_2^2+2
   \eta ^2 e^{x_1} ((x_1-1) x_2+x_3)+e^{2 x_1}\right)^2}.$$
As the Killing vector $\cJ$ vanishes, the \usugra s are satisfied with the dilaton
$$
\Phi=x_1 - \frac{1}{4} \ln\left(\eta ^2 x_2^2+2 \eta ^2 e^{x_1}((x_1-1) x_2+x_3)+e^{2 x_1}\right)^2.
$$

\subsubsection{Pluralities of $(B5+A_1|A_4)$}\label{sec:B5}

The semi-Abelian \mt\ $(B5+A_1|A_4)$ is given by commutation relations
$$
[T_1, T_2]=- T_2, \quad [T_1, T_3]=- T_3.
$$
To obtain flat models solving \eqref{betaG}--\eqref{betaPhi} we choose $E_0$ such that the metric has the form
$$
\cF(x) = \cG(x) = \left(
\begin{array}{cccc}
 E_{11} & E_{12} e^{-x_1} &
   E_{13} e^{-x_1} & E_{14} \\
 E_{12} e^{-x_1} & E_{22} e^{-2
   x_1} & 0 & 0 \\
 E_{13} e^{-x_1} & 0 & 0 & 0 \\
 E_{14} & 0 & 0 & E_{44} \\
\end{array}
\right).$$
Setting $E_{22} = E_{44} = E_{13} = 1$ and other $E_{ij} = 0$ we get flat \bkg\ with vanishing torsion. With $\varphi = x_1$ dilaton vanishes as well and the flux \eqref{FM} has components 
\begin{equation}\label{FA_5}
\mathcal F_A=(2,0,0,0,0,0,0,0).
\end{equation}

As we show in the Appendix, several decompositions of the \dd\ are related by \ybdn s. However, the only plural model with non-vanishing scalar curvature is obtained on the \mt\ $(B5+A_1|B5+A_1; \operatorname{P22})$, whose $\eta$-parametrized form
$$
[T_1, T_2]=- T_2, \quad [T_1, T_3]=- T_3, \quad [\widetilde{T}^2, \widetilde{T}^4]= \eta\widetilde{T}^2, \quad [\widetilde{T}^3, \widetilde{T}^4]= \eta\widetilde{T}^3
$$
is related to $(B5+A_1|A_4)$ by matrix \eqref{izoB4A4B4B4} that corresponds to homogeneous \ybdn\ with unimodular $R$ matrix. The plural metric and $\cB$-field can be calculated from
$$
\cF(x) = \frac{1}{\Delta}\left(
\begin{array}{cccc}
 -\eta ^2 x_3^2 & -\eta ^2 e^{-x_1} x_2 x_3 & \eta ^2 e^{-x_1} x_2^2+\eta ^2 x_3+e^{x_1} & -\eta  e^{x_1} x_3 \\
 -\eta ^2 e^{-x_1} x_2 x_3 & 2 \eta ^2 e^{-x_1} x_3+1 & -\eta ^2 e^{-x_1} x_2 & -\eta  x_2 \\
 \eta ^2 e^{-x_1} x_2^2+\eta ^2 x_3+e^{x_1} & -\eta ^2 e^{-x_1} x_2 & -\eta ^2 & -\eta  e^{x_1} \\
 \eta  e^{x_1} x_3 & \eta  x_2 & \eta  e^{x_1} & e^{2 x_1} \\
\end{array}
\right)
$$
where $\Delta = \eta ^2 x_2^2+2 \eta ^2 e^{x_1} x_3+e^{2 x_1}$.
The \bkg\ has non-trivial torsion and scalar curvature
$$
\cR=\frac{2 \eta ^2 e^{2 x_1} \left(-2 \eta ^2 x_2^2-4 \eta ^2 e^{x_1} x_3+5 e^{2 x_1}\right)}{\left(\eta ^2 x_2^2+2 \eta ^2 e^{x_1} x_3+e^{2 x_1}\right)^2}.
$$
Together with the dilaton
$$
\Phi=x_1 - \frac{1}{4} \ln \left(\eta ^2 x_2^2+2 \eta ^2 e^{x_1}x_3+e^{2 x_1}\right)^2
$$
it satisfies the \usugra s.

\subsubsection{Pluralities of $(B6_0+A_1|A_4)$}

Let us now focus on Manin triple $(B6_0+A_1|A_4)$. In Ref. \cite{snohla:DD} it was recognized that $(B6_0|A_3)$ belongs to the same 3+3-dimensional \dd\ as $(B5|A_3)$. As we show in the Appendix, there are much more possibilities for plurality \tfn s if we consider 4+4-dimensional \dd\ 4. Flat models on semi-Abelian \mt\ $(B6_0+A_1|A_4)$ with commutation relations
$$
[T_1, T_3]= T_2, \quad [T_2, T_3]= T_1
$$
are obtained from the constant matrix $E_0$
$$
E_0=\left(
\begin{array}{cccc}
 -E_{22} & 0 & E_{13} & 0 \\
 0 & E_{22} & E_{23} & 0 \\
 E_{13} & E_{23} & E_{33} & E_{34} \\
 0 & 0 & E_{34} & E_{44} \\
\end{array}
\right).$$
With $E_{22} = E_{33} = E_{44} = 1$ and the rest of the constants zero we get  Minkowski metric in the form
\begin{equation*}\label{mtziniB60}
\cF(x) = \cG(x)=\left(
\begin{array}{cccc}
 -1 & 0 & -x_2 & 0 \\
 0 & 1 & x_1 & 0 \\
 -x_2 & x_1 & x_1^2-x_2^2+1 & 0 \\
 0 & 0 & 0 & 1 \\
\end{array}
\right).
\end{equation*}
Both $\varphi$ and dilaton $\Phi$ vanish.

Plural solution of \sugra s with nonzero scalar curvature and torsion is obtained as \ybdn\ $(B6_0+A_1|A_4) \izo (B6_0+A_1|B6_0+A_1; \operatorname{P2})$ on \mt\ with commutation relations
$$
[T_1, T_3]= T_2, \quad [T_2, T_3]= T_1, \quad [\widetilde{T}^1, \widetilde{T}^4]= \eta \widetilde{T}^2, \quad [\widetilde{T}^2, \widetilde{T}^4]= \eta\widetilde{T}^1.
$$
The deformation is generated by unimodular $R$-matrix and
\begin{equation*}
M = \left(
\begin{array}{cccccccc}
 1 & 0 & 0 & 0 & 0 & 0 & 0 & 0 \\
 0 & 1 & 0 & 0 & 0 & 0 & 0 & 0 \\
 0 & 0 & 1 & 0 & 0 & 0 & 0 & 0 \\
 0 & 0 & 0 & 1 & 0 & 0 & 0 & 0 \\
 0 & 0 & 0 & 0 & 1 & 0 & 0 & 0 \\
 0 & 0 & 0 & 0 & 0 & 1 & 0 & 0 \\
 0 & 0 & 0 & \eta  & 0 & 0 & 1 & 0 \\
 0 & 0 & -\eta  & 0 & 0 & 0 & 0 & 1 \\
\end{array}
\right).
\end{equation*}
The resulting \bkg\ tensor with scalar curvature
$$
\cR = \frac{2 \eta^2 \left(\left(-2 x_1^2+2 x_2^2+5\right) \eta ^2+5\right)}{\left(\left(x_1^2-x_2^2+1\right) \eta ^2+1\right)^2}
$$
reads
$$\cf(x)=
\left(
\begin{array}{cccc}
 -\frac{\eta ^2 \left(x_1^2+1\right)+1}{\eta ^2 \left(x_1^2-x_2^2+1\right)+1} & \frac{\eta ^2 x_1 x_2}{\eta ^2 \left(x_1^2-x_2^2+1\right)+1} & -\frac{x_2}{\eta ^2 \left(x_1^2-x_2^2+1\right)+1} & \frac{\eta  x_2}{\eta ^2 \left(x_1^2-x_2^2+1\right)+1} \\
 \frac{\eta ^2 x_1 x_2}{\eta ^2 \left(x_1^2-x_2^2+1\right)+1} & \frac{1-\eta ^2 \left(x_2^2-1\right)}{\eta ^2 \left(x_1^2-x_2^2+1\right)+1} & \frac{x_1}{\eta ^2 \left(x_1^2-x_2^2+1\right)+1} & -\frac{\eta  x_1}{\eta ^2 \left(x_1^2-x_2^2+1\right)+1} \\
 -\frac{x_2}{\eta ^2 \left(x_1^2-x_2^2+1\right)+1} & \frac{x_1}{\eta ^2 \left(x_1^2-x_2^2+1\right)+1} & \frac{x_1^2-x_2^2+1}{\eta ^2 \left(x_1^2-x_2^2+1\right)+1} & \frac{\eta  \left(-x_1^2+x_2^2-1\right)}{\eta ^2 \left(x_1^2-x_2^2+1\right)+1} \\
 -\frac{\eta  x_2}{\eta ^2 \left(x_1^2-x_2^2+1\right)+1} & \frac{\eta  x_1}{\eta ^2 \left(x_1^2-x_2^2+1\right)+1} & \frac{\eta  \left(x_1^2-x_2^2+1\right)}{\eta ^2 \left(x_1^2-x_2^2+1\right)+1} & \frac{1}{\eta ^2 \left(x_1^2-x_2^2+1\right)+1} \\
\end{array}
\right),
$$
and together with dilaton
$$
\Phi = -\frac{1}{4} \ln \left(\left(x_1^2-x_2^2+1\right) \eta ^2+1\right)^2
$$
it satisfies the \sugra s.

It seems worth noting that while the \ybdn\ $(B6_0+A_1|A_4) \izo (B6_0+A_1|B2+A_1;\operatorname{P11})$ gives the flat \bkg, its dual on \mt\ $(B2+A_1;\operatorname{P11}|B6_0+A_1)$ leads to a \bkg\ with non-vanishing scalar curvature and torsion found in the Ref. \cite{hlapet:cqgrav} as non-Abelian dual of the flat metric. Plurality \tfn\ $(B6_0+A_1|A_4) \izo (B6_0+A_1; \operatorname{P2}| B6_0+A_1)$ provides the same results up to irrelevant constant terms in the $\cB$-field.

\subsubsection{Pluralities of $(B7_0+A_1|A_4)$} 

Flat models for \mt\ $(B7_0+A_1|A_4)$ with the commutation relations
$$
[T_1, T_3]=- T_2, \quad [T_2, T_3]= T_1
$$
are given by 
$$
E_0=\left(
\begin{array}{cccc}
E_{22} & 0 & E_{13} & 0 \\
 0 & E_{22} & E_{23} & 0 \\
 E_{13} & E_{23} & E_{33} & E_{34} \\
 0 & 0 & E_{34} & E_{44} \\
\end{array}
\right).
$$
Choosing $\varphi=0$ and $E_{22} = E_{44} = E_{34} = 1$ as the only nonzero components of $E_0$ we get vanishing dilaton $\Phi$ and flat Minkowski metric
\begin{equation*}\label{mtziniB70}
\cF (x) = \cG(x)=\left(
\begin{array}{cccc}
 1 & 0 & x_2 & 0 \\
 0 & 1 & -x_1 & 0 \\
 x_2 & -x_1 & x_1^2+x_2^2 & 1 \\
 0 & 0 & 1 & 1 \\
\end{array}
\right).
\end{equation*}
The \ybdn\ $(B7_0+A_1|A_4) \izo (B7_0+A_1|B7_0+A_1; \operatorname{P2}) $  is generated by unimodular $R$-matrix giving\footnote{In this case it is not be possible to perform plurality transformation with $\eta=1$ as then $\det \wh E_0= \frac{1}{\eta^2-1}$.}
\begin{equation}\label{M70}
M = \left(
\begin{array}{cccccccc}
 1 & 0 & 0 & 0 & 0 & 0 & 0 & 0 \\
 0 & 1 & 0 & 0 & 0 & 0 & 0 & 0 \\
 0 & 0 & 1 & 0 & 0 & 0 & 0 & 0 \\
 0 & 0 & 0 & 1 & 0 & 0 & 0 & 0 \\
 0 & 0 & 0 & 0 & 1 & 0 & 0 & 0 \\
 0 & 0 & 0 & 0 & 0 & 1 & 0 & 0 \\
 0 & 0 & 0 & -\eta  & 0 & 0 & 1 & 0 \\
 0 & 0 & \eta  & 0 & 0 & 0 & 0 & 1 \\
\end{array}\right).
\end{equation}

The resulting model on \mt\ $(B7_0+A_1|B7_0+A_1; \operatorname{P2})$ with commutation relations
$$
[T_1, T_3]=- T_2, \quad [T_2, T_3]= T_1, \quad [\widetilde{T}^1, \widetilde{T}^4]=-\eta \widetilde{T}^2, \quad [\widetilde{T}^2, \widetilde{T}^4]= \widetilde{T}^1
$$
is given by tensor field
$$
\cF(x) = 
\left(
\begin{array}{cccc}
 \frac{\eta ^2 \left(x_1^2-1\right)+1}{\eta ^2 \left(x_1^2+x_2^2-1\right)+1} & \frac{\eta ^2 x_1 x_2}{\eta ^2 \left(x_1^2+x_2^2-1\right)+1} & \frac{(\eta +1) x_2}{\eta ^2 \left(x_1^2+x_2^2-1\right)+1} & \frac{\eta  x_2}{\eta ^2 \left(x_1^2+x_2^2-1\right)+1} \\
 \frac{\eta ^2 x_1 x_2}{\eta ^2 \left(x_1^2+x_2^2-1\right)+1} & \frac{\eta ^2 \left(x_2^2-1\right)+1}{\eta ^2 \left(x_1^2+x_2^2-1\right)+1} & -\frac{(\eta +1) x_1}{\eta ^2 \left(x_1^2+x_2^2-1\right)+1} & -\frac{\eta  x_1}{\eta ^2 \left(x_1^2+x_2^2-1\right)+1} \\
 -\frac{(\eta -1) x_2}{\eta ^2 \left(x_1^2+x_2^2-1\right)+1} & \frac{(\eta -1) x_1}{\eta ^2 \left(x_1^2+x_2^2-1\right)+1} & \frac{x_1^2+x_2^2}{\eta ^2 \left(x_1^2+x_2^2-1\right)+1} & \frac{\eta  \left(x_1^2+x_2^2-1\right)+1}{\eta ^2 \left(x_1^2+x_2^2-1\right)+1} \\
 -\frac{\eta  x_2}{\eta ^2 \left(x_1^2+x_2^2-1\right)+1} & \frac{\eta  x_1}{\eta ^2 \left(x_1^2+x_2^2-1\right)+1} & \frac{1-\eta  \left(x_1^2+x_2^2-1\right)}{\eta ^2 \left(x_1^2+x_2^2-1\right)+1} & \frac{1}{\eta ^2 \left(x_1^2+x_2^2-1\right)+1} \\
\end{array}
\right)
$$
with
$$
\cR =\frac{2 \eta ^2 \left(\eta ^2 \left(2 x_1^2+2 x_2^2+5\right)-5\right)}{\left(\left(x_1^2+x_2^2-1\right) \eta ^2+1\right)^2}
$$
and non-trivial torsion $H$. Together with dilaton
$$
\Phi = -\frac{1}{4} \ln \left(\eta ^2 \left(x_1^2+x_2^2-1\right)+1\right)^2
$$
it satisfies \sugra s.

Similarly to the previous section we obtain \bkg s with non-trivial scalar curvature and torsion also as duals of the \ybdn s, i.e. on \mt s $(B2+A_1;\operatorname{P11}|B7_0+A_1)$ and $(B7_0+A_1;\operatorname{P2}|B7_0+A_1)$. These solutions of \sugra s were found in Ref. \cite{hlapet:cqgrav} as non-Abelian duals of the flat metric.

\subsection{Non-unimodular \yb deformations}\label{problematicke}

$R$-matrices for the decomposable four-dimensional algebras investigated above are unimodular, i.e. they satisfy the condition $R^{st}f_{st}{}^j =0$ for all $j$. This is not true for $R$-matrices appearing in transformations between \mt s containing indecomposable algebras. In this section we investigate non-unimodular \ybdn s to see whether they give solutions of \gsugra s.

\subsubsection{Pluralities of $(s_{4,11}|A_4)$}\label{sec:4_11}

The structure of semi-Abelian \mt\ $(s_{4,11}|A_4)$ is given by commutation relations of algebra $\mathfrak{g}=s_{4,11}$ as
$$
[T_1, T_4]=- T_1, \quad [T_2, T_3]= T_1, \quad [T_2, T_4]=- T_2.
$$
Flat models on the Lie group corresponding to Lie algebra $s_{4,11}$ can be constructed using constant matrix
\begin{equation}\label{E0411}
E_0=\left(
\begin{array}{cccc}
 0 & 0 & 0 & E_{14} \\
 0 & E_{22} & E_{14} & E_{24} \\
 0 & E_{14} & E_{33} & E_{34} \\
 E_{14} & E_{24} & E_{34} &
   E_{44} \\
\end{array}
\right).
\end{equation}
With $E_{22} = E_{14} = 1, E_{33} = 2$ and $E_{ij}= 0$ otherwise, we get flat metric
\begin{equation*}\label{metric_flat_s411}
\cF(x)=\cG(x)=\left(
\begin{array}{cccc}
 0 & 0 & 0 & 1 \\
 0 & 1 & 1 & -x_2 \\
 0 & 1 & 2 & 0 \\
 1 & -x_2 & 0 & x_2^2-2 x_1 \\
\end{array}
\right).
\end{equation*}
For $\varphi=-x_4$ the dilaton $\Phi$ vanishes, the \sugra s are trivially satisfied, and the flux \eqref{FM} has components 
$$
\mathcal F_A=(0,0,0,-2,0,0,0,0).
$$

The structure of the \dd\ 6 containing $(s_{4,11}|A_4)$ is rich, and various decompositions into standard \mt s exist including plurality between semi-Abelian \mt s $(s_{4,11}|A_4)  \cong (s_{4,6}|A_4)  $. We summarize full results in the Appendix. The \mt s containing $\cg=s_{4,11}$ related by \ybdn s are
$$
(s_{4,11}|A_4) \cong (s_{4,11}|B2+A_1; \operatorname{P10}) \cong (s_{4,11}|B2+A_1; \operatorname{P12}) \cong (s_{4,11}|s_{4,6}; \operatorname{P11}) \cong (s_{4,11}|s_{4,11}; \operatorname{P24}).
$$

The only solution of the \sugra s with nontrivial scalar curvature follows from the \tfn\ $(s_{4,11}|A_4)\izo (s_{4,11}|s_{4,6}; \operatorname{P11})$
given by  
$$M=\left(
\begin{array}{cccccccc}
 1 & 0 & 0 & 0 & 0 & 0 & 0 & 0 \\
 0 & 1 & 0 & 0 & 0 & 0 & 0 & 0 \\
 0 & 0 & 1 & 0 & 0 & 0 & 0 & 0 \\
 0 & 0 & 0 & 1 & 0 & 0 & 0 & 0 \\
 0 & 0 & 0 & 0 & 1 & 0 & 0 & 0 \\
 0 & 0 & 0 & -\eta & 0 & 1 & 0 & 0 \\
 0 & 0 & 0 & 0 & 0 & 0 & 1 & 0 \\
 0 & \eta & 0 & 0 & 0 & 0 & 0 & 1 \\
\end{array}
\right).$$
The corresponding $R$-matrix is non-unimodular with
$$
R^{st}f_{st}{}^c =(0,2\,\eta,0,0).
$$
For the \mt\ $(s_{4,11}|s_{4,6}; \operatorname{P11})$ with commutation relations
\begin{align*}
[T_1, T_4] &=- T_1, & [T_2, T_3] &= T_1, & [T_2, T_4] &=- T_2,\\ 
[\widetilde{T}^1, \widetilde{T}^2] &= -\eta \widetilde{T}^1, & [\widetilde{T}^1, \widetilde{T}^4] &= \eta\widetilde{T}^3, & [\widetilde{T}^2, \widetilde{T}^4] &=-\eta \widetilde{T}^4
\end{align*}
the \pltp y formulas yield rather extensive \bkg\ whose form is given by inverse of the matrix
\begin{equation*}\label{F411}
\cf\-1(x) =\left(
\begin{array}{cccc}
 x_2^2+2 x_1 & 2 x_2 & -x_2 & \eta\,e^{x_4}
   x_3+1 \\
 2 x_2 & 2 & -1 & -\eta\,e^{x_4} \\
 -x_2 & -1 & 1 & 0 \\
 1-\eta\,e^{x_4} x_3 & \eta\,e^{x_4} & 0 & 0 \\
\end{array}
\right).
\end{equation*} 
The \bkg\ has non-trivial torsion and scalar curvature
$$
\cR = 
\frac{4 \eta ^2 e^{2 x_4} \left(3 \eta ^2 e^{2 x_4} (2 x_1+x_3 (2 x_2+x_3))+11\right)}{\left(\eta ^2 e^{2 x_4} (2 x_1+x_3 (2 x_2+x_3))-1\right)^2}.
$$
Focusing on the transformation of the dilaton we first have
$$
\wh{\mathcal{F}}_A=(0,0,0,-2,0,2\eta,0,0).
$$
From the expression
$$
\wh{\mathcal{F}}_M=(0,0,0,-2,-2 \eta  e^{x_4} x_3, 2 \eta  e^{x_4},0,0)
$$
and formula \eqref{rhs} we can then calculate the dilaton and vector field $\cJ$ that have the form
\begin{equation*}\label{dil411}
\Phi = -\frac{1}{4} \ln \left(\eta ^2 e^{2 x_4} (2 x_1+x_3 (2 x_2+x_3))-1\right)^2, \qquad \cJ=\left\{-\eta  e^{x_4} x_3,\eta  e^{x_4},0,0\right\}.
\end{equation*}
We can verify that $\cJ$ is a Killing vector and that the fields satisfy the \gsugra s. Importantly, as $\mathrm{d}X\neq 0$, these fields cannot by transformed to solutions of \usugra s by gauge shift \eqref{gauge tfn lambda}, and represent a true solution to \gsugra s.

\subsubsection{Pluralities of $(s_{4,3}^{1,1}|A_4)$}

For the second example of \ybdn\ given by non-unimodular $R$-matrix we shall consider semi-Abelian \mt\ $(s_{4,3}^{1,1}|A_4)$ with commutation relations
$$
[T_1, T_4]=- T_1, \quad [T_2, T_4]=- T_2, \quad [T_3, T_4]=- T_3,
$$
which is a special case in the two-parametric family of \mt s $(s_{4,3}^{a,b}|A_4)$, see \cite{HNP} for details. 

For $(s_{4,3}^{a,b}|A_4)$ the Lie algebra invariants depend on $a$ and $b$ and for different values of parameters one obviously gets different \dd s. Therefore, it is quite challenging to properly identify \mt s belonging to the same \dd\ in the presence of parameters. We focus on the deformations of \mt\ containing algebra $\cg = s_{4,3}^{1,1}$, as it is a subalgebra of the Poincaré algebra \cite{PWZ}.

Flat Minkowski metric  on the Lie group corresponding to Lie algebra $\cg = s_{4,3}^{1,1}$ and simplified $E_0$ \eqref{E0411} has the form
$$
\cF(x)=\cG(x)=\left(
\begin{array}{cccc}
 0 & 0 & 0 & 1 \\
 0 & 1 & 1 & -x_2-x_3 \\
 0 & 1 & 2 & -x_2-2 x_3 \\
 1 & -x_2-x_3 & -x_2-2 x_3 & x_2^2+2 x_3 x_2+2 x_3^2-2 x_1 \\
\end{array}
\right).
$$
For $\varphi=-\frac{3}{2}x_4$ we get vanishing dilaton and the flux $\cf_A$ with components
$$\cF_A=(0,0,0,-3,0,0,0,0).$$

Plural \bkg\ with non-vanishing scalar curvature is obtained from the \tfn\ $(s_{4,3}^{1,1}|A_4)\izo(s_{4,3}^{1,1}|s_{4,3}^{1,-1};\operatorname{P2})$ given by matrix \eqref{M70}. The R-matrix is non-unimodular since
$$
R^{st}f_{st}{}^c =(0,0,2\eta,0).
$$ 

The deformed algebra $(s_{4,3}^{1,1}|s_{4,3}^{1,-1};\operatorname{P2})$ is given by commutation relations
\begin{align*}
[T_1, T_4] &=- T_1, & [T_2, T_4] &=- T_2, & [T_3, T_4] &=- T_3,\\
[\widetilde{T}^1, \widetilde{T}^3] &=- \eta\widetilde{T}^1, & [\widetilde{T}^2, \widetilde{T}^3] &=-\eta \widetilde{T}^2, & [\widetilde{T}^3, \widetilde{T}^4] &=-\eta \widetilde{T}^4.
\end{align*}
The rather extensive \bkg\ tensor can be found from the inverse of the matrix
\begin{equation}\label{F43a1b1}
\cf\-1(x) =\left(
\begin{array}{cccc}
 2 x_1 & x_2 & x_3 & 1 \\
 x_2 & 2 & -1 & 0 \\
 x_3 & -1 & 1 & -\eta \, e^{x_4} \\
 1 & 0 & \eta  \,e^{x_4} & 0 \\
\end{array}
\right)
\end{equation}
The \bkg\ has nonzero torsion and scalar curvature 
$$
\cR=\frac{8 \eta ^2 e^{2 x_4} \left(3 \eta ^2 e^{2 x_4} \left(4
   x_1-x_2^2\right)+11\right)}{\left(\eta ^2 e^{2 x_4} \left(4
   x_1-x_2^2\right)-1\right)^2}.
$$
To find the proper dilaton and $\cJ$ we first get
$$\wh{\mathcal F}_A=(0,0,0,-3,0,0,3\eta,0),$$ 
then
$$\wh{\mathcal F}_M=(0,0,0,-3,0,0,3 \eta  e^{x_4},0),$$
and finally, after the shift by $\half \bar f_b{}^{ba} \,\hat V_a^m$, compute the dilaton and Killing vector from components of $\wh{\mathcal F}_M$ as
\begin{equation}\label{dil43a1b1}
\Phi=-\frac{1}{4} \ln \left(\eta^2 e^{2 x_4} \left(4 x_1-x_2^2 \right)-1 \right)^2, \qquad \cJ=\left(0,0,\eta  e^{x_4},0\right).
\end{equation}
The \bkg\ \eqref{F43a1b1} together with dilaton and Killing vector $\cJ$ in \eqref{dil43a1b1} satisfy \gsugra s. Importantly, it is not possible to eliminate $\cJ$ by gauge \tfn\ since the one-form $X$ is not closed.

\subsection{Another solution of \gsugra s}

In Section \ref{sec:B5} we discussed \ybdn s of the flat model constructed on \mt\ $(B5+A_1|A_4)$. The structure of the \dd\ 4 is rich and beside the \ybdn s it is possible to consider other plurality transformations to obtain solutions of (generalized) \sugra s. Most of the plural \bkg s are flat or pp-waves. An interesting exception is the \tfn\ $(B5+A_1|A_4)\izo (B6_0+A_1|B6_0+A_1; \operatorname{P2})$ given by the matrix
\begin{equation}\label{M6060}
M=\left(
\begin{array}{cccccccc}
 0 & 0 & \frac{1}{2} & 0 & 0 & -\frac{1}{2} & 0 & 0 \\
 0 & 0 & \frac{1}{2} & 0 & 0 & \frac{1}{2} & 0 & 0 \\
 1 & 0 & 0 & 0 & 0 & 0 & 0 & 0 \\
 0 & 0 & 0 & 1 & 0 & 0 & 0 & 0 \\
 0 & -1 & 0 & 0 & 0 & 0 & 1 & 0 \\
 0 & 1 & 0 & 0 & 0 & 0 & 1 & 0 \\
 0 & 0 & 0 & 1 & 1 & 0 & 0 & 0 \\
 -1 & 0 & 0 & 0 & 0 & 0 & 0 & 1 \\
\end{array}
\right).
\end{equation}

For the \mt\  $(B6_0+A_1|B6_0+A_1; \operatorname{P2})$,  given by relations
$$
[T_1, T_3]= T_2, \quad [T_2, T_3]= T_1, \quad [\widetilde{T}^1, \widetilde{T}^4]= \widetilde{T}^2, \quad [\widetilde{T}^2, \widetilde{T}^4]= \widetilde{T}^1,
$$
the \pltp y yields \bkg\ with non-trivial torsion and scalar curvature
$$
\cR=\frac{16 \left(11 x_1^2-2 (11 x_2+6) x_1+11
   x_2^2-12 x_2+2\right)}{\left(x_1^2-2 (x_2-2)
   x_1+(x_2+2)^2\right)^2}.
$$
The metric and $\cB$-field can be found from the inverse of the matrix
$$
\cf\-1(x)=\left(
\begin{array}{cccc}
 1-2 x_2 & -x_1-x_2-1 & 1 & 0 \\
 -x_1-x_2-1 & 1-2 x_1 & 1 & 0 \\
 1 & 1 & 0 & 1 \\
 0 & 0 & -1 & 1 \\
\end{array}
\right).$$
Under the \tfn\ \eqref{M6060} the components of the flux \eqref{FA_5} become
$$\wh{\mathcal F}_A = (0,0,2,0,0,0,0,-2),$$
and since $\wh{\mathcal F}_M = \wh{\mathcal F}_A $, we get the dilaton and Killing vector
$$\Phi=x_3+\frac{1}{4} \ln \left(x_1^2-2 (x_2-2) x_1+(x_2+2)^2\right)^2, \qquad \cJ=   (0,0,0,-1).
$$
One-form $X$ is not closed in this case and we have obtained a solution to \gsugra s where the Killing vector cannot be eliminated by \eqref{gauge tfn lambda}. 

\section{pp-waves as non-unimodular deformations}\label{nonunimpp}

As follows from \eqref{alryPWZ}, there are two more subalgebras of the Poincare algebra, namely $s_{4,5}^{a,a}$ and $s_{4,5}^{a,0}$, both with $a>0$, which we should investigate. The corresponding semi-Abelian \mt s belong to different \dd s. In the Appendix we show that there are one-parametric families of \dd s depending on the value of parameter $a$. We were able to identify several \ybdn s relating the corresponding \mt s, but we did not obtain any \bkg s with non-vanishing scalar curvature using the plurality \tfn. Therefore, as the last example we shall discuss the \ybdn s generated by non-unimodular $R$-matrix that anyway lead to solutions of \usugra s, namely pp-wave \bkg s.

\subsection{Pluralities of $(s_{4,5}^{a,0}|A_4)$}

The commutation relations of the Lie algebra $\mathfrak{g}=s^{a,0}_{4,5}$ read
$$[T_1, T_4]=-a T_1, \quad [T_2, T_4]=T_3, \quad [T_3, T_4]=- T_2, \quad a > 0.$$
With a proper choice of $\varphi = -\frac{1}{2} a  x_4$ and $E_0$ we get vanishing dilaton, flat Minkowski metric
$$
\cF(x)=\cG(x)=\left(
\begin{array}{cccc}
 0 & 0 & 0 & -1 \\
 0 & 1 & 0 & -x_3 \\
 0 & 0 & 1 & x_2 \\
 -1 & -x_3 & x_2 & 2 a  x_1+x_2^2+x_3^2 \\
\end{array}
\right)
$$
and flux
$$\mathcal F_A = (0,0,0,-a,0,0,0,0).$$

The matrix $M$ of the \ybdn\ $(s_{4,5}^{a,0}|A_4) \izo (s_{4,5}^{a,0}|s_{4,5}^{a,0}; \operatorname{P22})$ has the form \eqref{izoB4A4B4B4} given by 
non-unimodular $R$-matrix since
$$R^{st}f_{st}{}^c =(-2 \eta\, a,0,0,0).$$
For the \mt\ $(s_{4,5}^{a,0}|s_{4,5}^{a,0}; \operatorname{P22})$ given by commutation relations
\begin{align*}
[T_1, T_4] &=-a T_1, & [T_2, T_4] &=T_3, & [T_3, T_4] &=- T_2, \\
[\widetilde{T}^1, \widetilde{T}^2] &=- \eta\widetilde{T}^3, & [\widetilde{T}^1, \widetilde{T}^3] &= \eta \widetilde{T}^2, & [\widetilde{T}^1, \widetilde{T}^4] &=a \eta \widetilde{T}^4
\end{align*}
we obtain pp-wave \bkg\ 
\begin{equation}\label{F45a0}
\cf(x)=\left( \begin{array}{cccc}
 0 & 0 & 0 & -\frac{1 }{\eta  e^{a  x_4}+1} \\
 0 & 1 & 0 & -\frac{x_3}{\eta  e^{a  x_4}+1} \\
 0 & 0 & 1 & \frac{x_2}{\eta  e^{a  x_4}+1} \\
 \frac{1 }{\eta  e^{a  x_4}-1} & \frac{x_3}{\eta  e^{a 
   x_4}-1} & \frac{x_2}{1-\eta  e^{a  x_4}} & \frac{2 a  
   x_1+x_2^2+x_3^2}{1-\eta ^2 e^{2 a  x_4}} \\
\end{array}
\right)
\end{equation}
with non-trivial torsion. The formula \eqref{rhs} gives correct dilaton and Killing vector $\cJ$ as
\begin{equation}\label{dil45a0}
\Phi=-\frac{1}{4} \ln\left(\eta ^2 e^{2 a  x_4}-1\right)^2, \qquad \cJ=\left(-a  \eta e^{a x_4},0,0,0\right).
\end{equation}
Since the one-form $X$ is closed, we can use the gauge \tfn\ \eqref{gauge tfn lambda} to reduce the solution to a solution of \usugra s.

On the other hand, one can check that due to the nontrivial torsion the \sugra s are also satisfied with both dilaton and $\cJ$ vanishing. This ambiguity follows from the fact that the one-form $X$ that solves the (standard) \sugra s  can be chosen as
$$X= \frac{C\, a\, \eta\, e^{a x_4}}{1-\eta ^2 e^{2 a x_4}} \mathrm{d}x_4,$$
where $C$ is arbitrary including zero. One gets very similar results for pluralities of $(s_{4,5}^{a,a}|A_4)$.

\subsection{Plurality of $(s_{4,11}|A_4)$}

Beside the plurality investigated in Sec. \ref{problematicke}, there is one more \mt\ that is \dd\ isomorphic to $(s_{4,11}|A_4)$ via non-unimodular $R$-matrix
$$R=\left(\begin{array}{cccc}
0 & 0 & 0 & 1
\\
 0 & 0 & -1 & 0
\\
 0 & 1 & 0 & 0
\\
 -1 & 0 & 0 & 0
\end{array}\right), \quad R^{st}f_{st}{}^c =(-4,0,0,0),
$$
namely $(s_{4,11}|s_{4,11}; \operatorname{P24})$. To evade the singularity of $\wh{E_0}$ we choose $E_{33} = \frac{5}{2}, E_{14} = E_{22} = \frac{1}{2}$ and $E_{ij}=0$ for other constants in \eqref{E0411}.

The pp-wave \bkg\ can be calculated as the inverse of
  \begin{equation}\label{F411pp}
\cf\-1 (x)= \left(
\begin{array}{cccc}
 \frac{1}{2} \left(x_2^2+8 x_1\right) & \frac{5 x_2}{2} & \eta  e^{x_4} x_3-\frac{x_2}{2} &
   \eta  e^{x_4}+2 \\
 \frac{5 x_2}{2} & \frac{5}{2} & -\eta e^{x_4}-\frac{1}{2} & 0 \\
 -\eta  e^{x_4} x_3-\frac{x_2}{2} & \eta  e^{x_4}-\frac{1}{2} & \frac{1}{2} & 0 \\
 2-\eta  e^{x_4} & 0 & 0 & 0 \\
\end{array}
\right).
\end{equation}
Dilaton
\begin{equation}\label{dil4511p}
\Phi= - \frac{1}{4} \ln \left(-\eta ^4 e^{4 x_4}+3 \eta ^2 e^{2 x_4}+4\right)^2
\end{equation}
and the Killing vector
$$\cJ=\left\{-2 \eta  e^{x_4},0,0,0\right\}$$
obtained from the formulas \eqref{eq:dilaton-JL}, \eqref{rhs} yield closed one-form $X$ so that the \usugra s are satisfied.

\section{Conclusions}
In this work we have investigated solutions of (generalized) \sugra s obtained by the \pltp y  of the flat Minkowski metric in the dimension four. To apply the \pltp y as a solution-generating technique, we used the classification of 4+4-dimensional standard \mt s presented recently in Ref. \cite{HNP}. Considering semi-Abelian \mt s $(\mathfrak g|A_4)$ containing subalgebras $\cg$ of the Poincaré algebra we have identified which standard \mt s form the same \dd, and found the particular mappings between their generators. The results are summarized in the Appendix.

Many 4+4-dimensional standard \mt s turn out to be of the \yb type given by \eqref{ybtfn}. In such cases the $O(D,D)$ transformation between \mt s can be chosen to have the form
\begin{equation*}
 M=\left(\begin{array}{cc} \unit & 0\\  
 R & \unit \\ \end{array}\right)
\end{equation*}
given by $R$-matrix satisfying homogeneous \yb equation \eqref{cybe2}. Introducing deformation parameter $\eta$ we understand the corresponding \pltp ies as \ybdn s of the flat \sm.

Most of the plural \bkg s are either flat or pp-waves with or without torsion satisfying \usugra s \eqref{betaG}--\eqref{betaPhi} with $\cJ = 0$. Being interested in solutions with non-vanishing $\cJ$, we have focused mainly on the \bkg s with non-vanishing scalar curvature.

The (non-)unimodularity of the $R$-matrix defining the \ybdn\ is closely related to the type of the \sugra s that the resulting \bkg s satisfy. The \bkg s obtained in Sec. \ref{YangBaxter} by \tfn s with unimodular $R$-matrices solve the \usugra s. For non-unimodular $R$-matrices we obtain solutions of \gsugra s with non-trivial Killing vectors $\cJ$ in the Section \ref{problematicke}. However, when the corresponding one-form $X$ is closed, one may use a gauge transformation to eliminate $\cJ$. Examples of these \yb deformations that solve \usugra s even though they were generated by non-unimodular $R$-matrices are the pp-waves presented in Sec. \ref{nonunimpp}. Besides that, the examples in the Sec. \ref{nonunimpp} show that the one-forms $X$  solving the \sugra s need not be unique. 

\section{Appendix: Examples of the  Drinfeld doubles}

Bellow  we give the list of  Drinfeld doubles -- sets of standard \mt s with equal invariants that were used for finding the solutions of the \sugra s. A Manin triple $(\cg, \tcg)$ is called standard and is denoted by $(\cg \mid \tcg; \operatorname{Pk}, \epsilon)$ if the structure constants of $\cg$ are taken directly from the list of four-dimensional Lie algebras, while the structure constants of $\tcg$ are obtained by applying the $k$-th permutation and a scaling factor $\epsilon = \pm 1$ to its basis elements $\{\widetilde T^1, \widetilde T^2, \widetilde T^3, \widetilde T^4\}$. For more details, notation of the \mt s, and their Lie products see \cite{HNP}. Semi-Abelian Manin triples are abbreviated as $(\cg \mid A_4)$.

The invariants of the Lie algebra $\cd$ of the \dd\ that we calculate are the dimensions of derived series
$$
\cd^{0} = \cd, \qquad \cd^{k+1} = [ \cd^{k}, \cd^{k} ],
$$
lower central series
$$
\cd^{0} = \cd, \qquad \cd^{k+1} = [ \cd^{k}, \cd],
$$
and algebra of derivations $Der \cd$. Besides that we compute the signature of the Killing form
\begin{equation}
K(x,y)=\operatorname{Tr}(\operatorname{ad}(x)\operatorname{ad}(y)),
\end{equation}
whose elements can be calculated from structure constants of $\cd$ as
\begin{equation}
K_{IJ} = X_{IM}{}^{N} X_{JN}{}^{M}.
\end{equation}
We list the signatures of the Killing form according to multiplicity or number of positive/negative/zero eigenvalues in the following way
$$
\{\text{\# positive}, \text{\# zero}, \text{\# negative}\}.
$$
An isomorphism of Drinfeld doubles $(\cg|\widetilde{\cg}) \izo (\cg' |\widetilde \cg')$ means that their corresponding bases $(T,\widetilde T)$ and $(T',\widetilde T')$ are related by
\begin{equation}
\left(\begin{array}{c} T'\\ \widetilde T' \end{array} \right) = M\left(\begin{array}{c} T\\ \widetilde{T}\end{array}\right)=\left(\begin{array}{cc} P & Q\\ R & S \\ \end{array}\right)\left(\begin{array}{c} T\\ \widetilde{T}\end{array}\right).
\end{equation}
For \ybdn s the matrices $M$ are given by $P=S=\bf{1}_4$, $Q=\bf{0}_4$, and we list only the $R$-matrices (or $Q$ if $R=\bf{0}_4$).

\renewcommand{\arraystretch}{1.2}
\subsection*{Drinfeld double \arabic{DDCounter}}
\addtocounter{DDCounter}{1}
\begin{tabular}{|c|c|c|c|l|}
\hline
Derived  & Lower central & Algebra of  &  Killing form & Manin triples \\
series  & series & derivations & signature  &\\
\hline
\multirow{4}{*}{$\{8,3,1,0\}$} & \multirow{4}{*}{$\{8,3\}$} & \multirow{4}{*}{29} & \multirow{4}{*}{$\{1,7,0\}$}
  & $(s_{2,1}+A_2|A_4)$\\
 & & & & $(s_{2,1}+A_2|B2+A_1; \operatorname{P1})$\\
 & & & & $(s_{2,1}+A_2|s_{2,1}+A_2; \operatorname{P7})$\\
 & & & & $(s_{2,1}+A_2|s_{2,1}+A_2; \operatorname{P15})$\\
\hline
\end{tabular}\\[8pt]
\begin{tabular}{l@{\hspace{0.5cm}}l}
$(s_{2,1}+A_2|A_4) \izo (s_{2,1}+A_2|B2+A_1; \operatorname{P1}) $ & $
R=\left(\begin{array}{cccc}
0 & 0 & 0 & 0
\\
 0 & 0 & 1 & 0
\\
 0 & -1 & 0 & 0
\\
 0 & 0 & 0 & 0
\end{array}\right)$\\
$(s_{2,1}+A_2|A_4) \izo (s_{2,1}+A_2|s_{2,1}+A_2; \operatorname{P7}) $ & $
R=\left(\begin{array}{cccc}
0 & -1 & 0 & 0
\\
 1 & 0 & 0 & 0
\\
 0 & 0 & 0 & 0
\\
 0 & 0 & 0 & 0
\end{array}\right)$\\
$(s_{2,1}+A_2|A_4) \izo (s_{2,1}+A_2|s_{2,1}+A_2; \operatorname{P15}) $ & $
R=\left(\begin{array}{cccc}
0 & 0 & 1 & 0
\\
 0 & 0 & 0 & 0
\\
 -1 & 0 & 0 & 0
\\
 0 & 0 & 0 & 0
\end{array}\right)$
\end{tabular}

\subsection*{Drinfeld double \arabic{DDCounter}}\addtocounter{DDCounter}{1}
\begin{tabular}{|c|c|c|c|l|}
\hline
Derived  & Lower central & Algebra of  &  Killing form & Manin triples \\
series  & series & derivations & signature  &\\
\hline
\multirow{5}{*}{$\{8,5,1,0\}$} & \multirow{5}{*}{$\{8,5,3\}$} & \multirow{5}{*}{18} & \multirow{5}{*}{$\{1,7,0\}$}
 & $(s_{2,1}+A_2|B2+A_1; \operatorname{P5})$\\
 & & & & $(s_{4,1}|A_4)$\\
 & & & & $(s_{4,1}|B2+A_1; \operatorname{P12})$\\
 & & & & $(s_{4,1}|B2+A_1; \operatorname{P22})$\\
 & & & & $(s_{4,1}|s_{2,1}+A_2; \operatorname{P3})$\\
\hline
\end{tabular}\\[8pt]
\begin{tabular}{l@{\hspace{0.5cm}}l}
$(s_{4,1}|A_4) \izo (s_{2,1}+A_2|B2+A_1; \operatorname{P5}) $ & $
M=\left(\begin{array}{cccccccc}
0 & 0 & 0 & 1 & 0 & 0 & 0 & 0
\\
 0 & 0 & 1 & 0 & 0 & 0 & 0 & 0
\\
 0 & 0 & 0 & 0 & 0 & 1 & 0 & 0
\\
 1 & 0 & 0 & 0 & 0 & 0 & 0 & 0
\\
 0 & 0 & 0 & 0 & 0 & 0 & 0 & 1
\\
 0 & 0 & 0 & 0 & 0 & 0 & 1 & 0
\\
 0 & 1 & 0 & 0 & 0 & 0 & 0 & 0
\\
 0 & 0 & 0 & 0 & 1 & 0 & 0 & 0
\end{array}\right)
$\\
$(s_{4,1}|A_4) \izo (s_{4,1}|B2+A_1; \operatorname{P12}) $ & $
R=\left(\begin{array}{cccc}
0 & 0 & 1 & 0\\
 0 & 0 & 0 & 0\\
 -1 & 0 & 0 & 0\\
 0 & 0 & 0 & 0
\end{array}\right)
$\\
$(s_{4,1}|A_4) \izo (s_{4,1}|B2+A_1; \operatorname{P22}) $ & $
R=\left(\begin{array}{cccc}
0 & 0 & -1 & 0 \\
 0 & 0 & 1 & 0 \\
 1 & -1 & 0 & 0 \\
 0 & 0 & 0 & 0
\end{array}\right)
$\\
$(s_{4,1}|A_4) \izo (s_{4,1}|s_{2,1}+A_2; \operatorname{P3})$ & $
R=\left(\begin{array}{cccc}
0 & 0 & 0 & -1 \\
 0 & 0 & 0 & 0\\
 0 & 0 & 0 & 0\\
 1 & 0 & 0 & 0
\end{array}\right)$
\end{tabular}

\subsection*{Drinfeld double \arabic{DDCounter}}\label{ddB4}\addtocounter{DDCounter}{1}
\begin{tabular}{|c|c|c|c|l|}
\hline
Derived  & Lower central & Algebra of  &  Killing form & Manin triples \\
series  & series & derivations & signature  &\\
\hline
\multirow{6}{*}{$\{8,5,1,0\}$} & \multirow{6}{*}{$\{8,5\}$} & \multirow{6}{*}{16} & \multirow{6}{*}{$\{1,7,0\}$}
  & $(B4+A_1|A_4)$\\
 & & & & $(B4+A_1|B2+A_1; \operatorname{P1}, \epsilon)$\\
 & & & & $(B4+A_1|B2+A_1; \operatorname{P2})$\\
 & & & & $(B4+A_1|B2+A_1; \operatorname{P5})$\\
 & & & & $(B4+A_1|B4+A_1; \operatorname{P24})$\\
 & & & & $(B6_0+A_1|B2+A_1; \operatorname{P9})$\\
\hline
\end{tabular}\\[8pt]
\begin{tabular}{l@{\hspace{0.5cm}}l}
$(B4+A_1|A_4) \izo (B4+A_1|B2+A_1; \operatorname{P1}, \epsilon) $ & $
R=\left(\begin{array}{cccc}
0 & 0 & 0 & 0
\\
 0 & 0 & -\frac{\epsilon}{2} & 0
\\
 0 & \frac{\epsilon}{2} & 0 & 0
\\
 0 & 0 & 0 & 0
\end{array}\right)$\\
$(B4+A_1|A_4) \izo (B4+A_1|B2+A_1; \operatorname{P2}) $ & $
R=\left(\begin{array}{cccc}
0 & 0 & 0 & 0
\\
 0 & 0 & 0 & -1
\\
 0 & 0 & 0 & -1
\\
 0 & 1 & 1 & 0
\end{array}\right)$\\
$(B4+A_1|A_4) \izo (B4+A_1|B2+A_1; \operatorname{P5}) $ & $
R=\left(\begin{array}{cccc}
0 & 0 & 0 & 0
\\
 0 & 0 & 0 & 0
\\
 0 & 0 & 0 & -1
\\
 0 & 0 & 1 & 0
\end{array}\right)$\\
$(B4+A_1|A_4) \izo (B4+A_1|B4+A_1; \operatorname{P24}) $ & $
R=\left(\begin{array}{cccc}
0 & 0 & 0 & 1
\\
 0 & 0 & 0 & 0
\\
 0 & 0 & 0 & 0
\\
 -1 & 0 & 0 & 0
\end{array}\right)$\\
$(B4+A_1|A_4) \izo (B6_0+A_1|B2+A_1; \operatorname{P9}) $ & $
M=\left(\begin{array}{cccccccc}
0 & 0 & 1 & 0 & 0 & \frac{1}{2} & 0 & 0
\\
 0 & 0 & 1 & 0 & 0 & -\frac{1}{2} & 0 & 0
\\
 1 & 0 & 0 & 0 & 0 & 0 & 0 & 0
\\
 0 & 0 & 0 & 1 & 0 & 0 & 0 & 0
\\
 0 & 1 & 0 & 0 & 0 & 0 & \frac{1}{2} & 0
\\
 0 & -1 & 0 & 0 & 0 & 0 & \frac{1}{2} & 0
\\
 0 & 0 & 0 & 0 & 1 & 0 & 0 & 0
\\
 0 & 0 & 0 & 0 & 0 & 0 & 0 & 1
\end{array}\right)$
\end{tabular}

\subsection*{Drinfeld double \arabic{DDCounter}\label{ddB5}} 
\addtocounter{DDCounter}{1}
\begin{tabular}{|c|c|c|c|l|}
\hline
Derived  & Lower central & Algebra of  &  Killing form & Manin triples \\
series  & series & derivations & signature  &\\
\hline
\multirow{7}{*}{$\{8,5,1,0\}$} & \multirow{7}{*}{$\{8,5\}$} & \multirow{7}{*}{18} & \multirow{7}{*}{$\{1,7,0\}$}
  & $(B5+A_1|A_4)$\\
 & & & & $(B5+A_1|B2+A_1; \operatorname{P1})$\\
 & & & & $(B5+A_1|B2+A_1; \operatorname{P2})$\\
 & & & & $(B5+A_1|B5+A_1; \operatorname{P22})$\\
 & & & & $(B6_0+A_1|A_4)$\\
 & & & & $(B6_0+A_1|B2+A_1; \operatorname{P11})$\\
 & & & & $(B6_0+A_1|B6_0+A_1; \operatorname{P2})$\\
\hline
\end{tabular}\\[8pt]
\begin{tabular}{l@{\hspace{0.5cm}}l}
$(B5+A_1|A_4) \izo (B5+A_1|B2+A_1; \operatorname{P1}) $ & $
R=\left(\begin{array}{cccc}
0 & 0 & 0 & 0
\\
 0 & 0 & -\frac{1}{2} & 0
\\
 0 & \frac{1}{2} & 0 & 0
\\
 0 & 0 & 0 & 0
\end{array}\right)$\\
$(B5+A_1|A_4) \izo (B5+A_1|B2+A_1; \operatorname{P2}) $ & $
R=\left(\begin{array}{cccc}
0 & 0 & 0 & 0
\\
 0 & 0 & 0 & -1
\\
 0 & 0 & 0 & 0
\\
 0 & 1 & 0 & 0
\end{array}\right)$\\
$(B5+A_1|A_4) \izo (B5+A_1|B5+A_1; \operatorname{P22}) $ & $
R=\left(\begin{array}{cccc}
0 & 0 & 0 & 1
\\
 0 & 0 & 0 & 0
\\
 0 & 0 & 0 & 0
\\
 -1 & 0 & 0 & 0
\end{array}\right)$\\
$(B6_0+A_1|A_4) \izo (B5+A_1|A_4)$ & $
M=\left(\begin{array}{cccccccc}
0 & 0 & 1 & 0 & 0 & 0 & 0 & 0
\\
 0 & 0 & 0 & 0 & -\frac{1}{2} & \frac{1}{2} & 0 & 0
\\
 1 & 1 & 0 & 0 & 0 & 0 & 0 & 0
\\
 0 & 0 & 0 & 1 & 0 & 0 & 0 & 0
\\
 0 & 0 & 0 & 0 & 0 & 0 & 1 & 0
\\
 -1 & 1 & 0 & 0 & 0 & 0 & 0 & 0
\\
 0 & 0 & 0 & 0 & \frac{1}{2} & \frac{1}{2} & 0 & 0
\\
 0 & 0 & 0 & 0 & 0 & 0 & 0 & 1
\end{array}\right)$\\
$(B6_0+A_1|A_4) \izo (B6_0+A_1|B2+A_1; \operatorname{P11}) $ & $
R=\left(\begin{array}{cccc}
0 & 0 & 0 & 0
\\
 0 & 0 & 0 & -1
\\
 0 & 0 & 0 & 0
\\
 0 & 1 & 0 & 0
\end{array}\right)$\\
$(B6_0+A_1|A_4) \izo (B6_0+A_1|B6_0+A_1; \operatorname{P2}) $ & $
R=\left(\begin{array}{cccc}
0 & 0 & 0 & 0
\\
 0 & 0 & 0 & 0
\\
 0 & 0 & 0 & 1
\\
 0 & 0 & -1 & 0
\end{array}\right)$
\end{tabular}

\subsection*{Drinfeld double \arabic{DDCounter}}\addtocounter{DDCounter}{1}
\begin{tabular}{|c|c|c|c|l|}
\hline
Derived  & Lower central & Algebra of  &  Killing form & Manin triples \\
series  & series & derivations & signature  &\\
\hline
\multirow{3}{*}{$\{8,5,1,0\}$} & \multirow{3}{*}{$\{8,5\}$} & \multirow{3}{*}{18} & \multirow{3}{*}{$\{0,7,1\}$}
  & $(B7_0+A_1|A_4)$\\
 & & & & $(B7_0+A_1|B2+A_1; \operatorname{P11})$\\
 & & & & $(B7_0+A_1|B7_0+A_1; \operatorname{P2})$\\
\hline
\end{tabular}\\[8pt]
\begin{tabular}{l@{\hspace{0.5cm}}l}
$(B7_0+A_1|A_4) \izo (B7_0+A_1|B2+A_1; \operatorname{P11}) $ & $
R=\left(\begin{array}{cccc}
0 & 0 & 0 & 0
\\
 0 & 0 & 0 & -1
\\
 0 & 0 & 0 & 0
\\
 0 & 1 & 0 & 0
\end{array}\right)$\\
$(B7_0+A_1|A_4) \izo (B7_0+A_1|B7_0+A_1; \operatorname{P2}) $ & $
R=\left(\begin{array}{cccc}
0 & 0 & 0 & 0
\\
 0 & 0 & 0 & 0
\\
 0 & 0 & 0 & -1
\\
 0 & 0 & 1 & 0
\end{array}\right)$
\end{tabular}

\subsection*{Drinfeld double \arabic{DDCounter}}\label{s411A4}\addtocounter{DDCounter}{1}
\begin{tabular}{|c|c|c|c|l|}
\hline
Derived  & Lower central & Algebra of  &  Killing form & Manin triples \\
series  & series & derivations & signature  &\\
\hline
\multirow{16}{*}{$\{8,6,2,0\}$} & \multirow{16}{*}{$\{8,6\}$} & \multirow{16}{*}{13} & \multirow{16}{*}{$\{1,7,0\}$}
  & $(B4+A_1|B2+A_1; \operatorname{P19})$\\
& & & & $(B5+A_1|B2+A_1; \operatorname{P19})$\\
& & & & $(B5+A_1|B4+A_1; \operatorname{P22})$\\
& & & & $(B6_0+A_1|B2+A_1; \operatorname{P10})$\\
& & & & $(n_{4,1}|B4+A_1; \operatorname{P1})$\\
& & & & $(n_{4,1}|B4+A_1; \operatorname{P2}, \epsilon)$\\
& & & & $(n_{4,1}|B5+A_1; \operatorname{P1})$\\
& & & & $(n_{4,1}|B5+A_1; \operatorname{P2})$\\
& & & & $(n_{4,1}|B6_0+A_1; \operatorname{P17})$\\
& & & & $(s_{4,6}|A_4)$\\
& & & & $(s_{4,6}|B2+A_1; \operatorname{P10})$\\
& & & & $(s_{4,11}|A_4)$\\
& & & & $(s_{4,11}|B2+A_1; \operatorname{P10})$\\
& & & & $(s_{4,11}|B2+A_1; \operatorname{P12})$\\
& & & & $(s_{4,11}|s_{4,6}; \operatorname{P11})$\\
& & & & $(s_{4,11}|s_{4,11}; \operatorname{P24})$\\
\hline
\end{tabular}\\[8pt]
\begin{tabular}{l@{\hspace{0.5cm}}l}
$(s_{4,11}|A_4)  \izo (B4+A_1|B2+A_1; \operatorname{P19}) $ & $
M=\left(\begin{array}{cccccccc}
0 & 0 & 1 & -1 & 0 & 0 & 0 & 0
\\
 0 & 1 & 0 & 0 & 0 & 0 & 0 & 0
\\
 -1 & 0 & 0 & 0 & 0 & 0 & 0 & 0
\\
 0 & 0 & 0 & 0 & 0 & 0 & 1 & 1
\\
 0 & 0 & 0 & 0 & 0 & 0 & 0 & -1
\\
 0 & 0 & 0 & 0 & 0 & 1 & 0 & 0
\\
 0 & 0 & 0 & 0 & -1 & 0 & 0 & 0
\\
 0 & 0 & 1 & 0 & 0 & 0 & 0 & 0
\end{array}\right)$\\
$(B4+A_1|B2+A_1; \operatorname{P19}) \izo (B5+A_1|B2+A_1; \operatorname{P19}) $ & $
Q=\left(\begin{array}{cccc}
0 & 0 & 0 & -1
\\
 0 & 0 & 0 & 0
\\
 0 & 0 & 0 & 0
\\
 1 & 0 & 0 & 0
\end{array}\right)$\\
$(s_{4,11}|A_4)  \izo (B5+A_1|B4+A_1; \operatorname{P22}) $ & $
M=\left(\begin{array}{cccccccc}
0 & 0 & 0 & -1 & 0 & 0 & 0 & 0
\\
 1 & 0 & 0 & 0 & 0 & 0 & 0 & 0
\\
 0 & 1 & 0 & 0 & 0 & 0 & 0 & 0
\\
 0 & 0 & 0 & 0 & 0 & 0 & 1 & 0
\\
 0 & 0 & 0 & 0 & 0 & 0 & 1 & -1
\\
 0 & 0 & 0 & 0 & 1 & 0 & 0 & 0
\\
 0 & 0 & 0 & 0 & 0 & 1 & 0 & 0
\\
 0 & 0 & 1 & 1 & 0 & 0 & 0 & 0
\end{array}\right)$\\
$(s_{4,11}|A_4)  \izo (B6_0+A_1|B2+A_1; \operatorname{P10}) $ & $
M=\left(\begin{array}{cccccccc}
\frac{1}{2} & 0 & 0 & 0 & 0 & 1 & 0 & 0
\\
 -\frac{1}{2} & 0 & 0 & 0 & 0 & 1 & 0 & 0
\\
 0 & 0 & 0 & 1 & 0 & 0 & 0 & 0
\\
 0 & 0 & 1 & 0 & 0 & 0 & 0 & 0
\\
 0 & \frac{1}{2} & 0 & 0 & 1 & 0 & 0 & 0
\\
 0 & \frac{1}{2} & 0 & 0 & -1 & 0 & 0 & 0
\\
 0 & 0 & 0 & 0 & 0 & 0 & 0 & 1
\\
 0 & 0 & 0 & 0 & 0 & 0 & 1 & 0
\end{array}\right)$\\
$(s_{4,11}|A_4) \izo (n_{4,1}|B4+A_1; \operatorname{P1}) $ & $
M=\left(\begin{array}{cccccccc}
0 & 0 & 0 & 0 & 0 & 0 & 0 & 1
\\
 1 & 0 & 0 & 0 & 0 & 0 & -1 & 1
\\
 1 & 1 & 0 & 0 & 0 & 0 & -1 & 0
\\
 -1 & 0 & 1 & 0 & 1 & 0 & 1 & 0
\\
 0 & 1 & 1 & 1 & 0 & 1 & -1 & 0
\\
 0 & 0 & 0 & 0 & 1 & -1 & 1 & 0
\\
 0 & 0 & 0 & 0 & 0 & 1 & 0 & -1
\\
 0 & 0 & 0 & 0 & 0 & 0 & 1 & -1
\end{array}\right)$\\
\end{tabular}\\
\begin{tabular}{l@{\hspace{0.5cm}}l}
$(s_{4,11}|A_4) \izo (n_{4,1}|B4+A_1; \operatorname{P2}, \epsilon) $ & $
M=\left(\begin{array}{cccccccc}
0 & 0 & 0 & 0 & 0 & 0 & 0 & \epsilon
\\
 \frac{\epsilon}{2} & 0 & 0 & 0 & 0 & \frac{-\epsilon}{2} & 0 & 0
\\
 0 & 0 & \frac{-\epsilon}{2} & 0 & 0 & 0 & 0 & 0
\\
 0 & 1 & 0 & 0 & 1 & 0 & 0 & 0
\\
 0 & 0 & \frac{1}{2} & \epsilon  & 0 & 0 & 0 & 0
\\
 0 & 0 & 0 & 0 & 2 \epsilon  & 0 & 0 & 0
\\
 0 & 0 & 0 & 0 & 0 & 0 & -2 \epsilon  & 1
\\
 0 & 0 & 0 & 0 & 0 & 1 & 0 & 0
\end{array}\right)$\\
$(n_{4,1}|B4+A_1; \operatorname{P1}) \izo (n_{4,1}|B5+A_1; \operatorname{P1}) $ & $
R=\left(\begin{array}{cccc}
0 & 0 & 0 & -1
\\
 0 & 0 & 0 & 0
\\
 0 & 0 & 0 & 0
\\
 1 & 0 & 0 & 0
\end{array}\right)$\\
$(s_{4,11}|A_4)  \izo (n_{4,1}|B5+A_1; \operatorname{P2})  $ & $
M=\left(\begin{array}{cccccccc}
0 & 0 & 0 & 0 & 0 & 0 & 0 & 1
\\
 \frac{1}{2} & 0 & 0 & 0 & 0 & 1 & 0 & 0
\\
 0 & 0 & 1 & 0 & 0 & 0 & 0 & 0
\\
 0 & -\frac{1}{2} & 0 & 0 & 1 & 0 & 0 & 0
\\
 0 & 0 & 0 & 1 & 0 & 0 & 0 & 0
\\
 0 & 0 & 0 & 0 & 2 & 0 & 0 & 0
\\
 0 & 0 & 0 & 0 & 0 & 0 & 1 & 0
\\
 0 & 0 & 0 & 0 & 0 & -2 & 0 & 0
\end{array}\right)
$\\
$(s_{4,11}|A_4)  \izo (n_{4,1}|B6_0+A_1; \operatorname{P17})  $ & $
M=\left(\begin{array}{cccccccc}
0 & 0 & 0 & 0 & 0 & 0 & 0 & 1
\\
 -1 & 0 & 0 & 0 & 0 & -\frac{1}{2} & 1 & 0
\\
 0 & 1 & 1 & 0 & \frac{1}{2} & 0 & 0 & 0
\\
 0 & 1 & 0 & 0 & -\frac{1}{2} & 0 & 0 & 0
\\
 0 & 0 & 0 & 1 & 0 & 0 & 0 & 0
\\
 0 & 0 & 1 & 0 & 0 & 0 & 0 & 0
\\
 1 & 0 & 0 & 0 & 0 & \frac{1}{2} & 0 & 0
\\
 -1 & 0 & 0 & 0 & 0 & \frac{1}{2} & 0 & 0
\end{array}\right)$\\
$(s_{4,11}|A_4)  \izo (s_{4,6}|A_4)  $ & $
M=\left(\begin{array}{cccccccc}
0 & 0 & 0 & 0 & 0 & 0 & 1 & 0
\\
 0 & 1 & 0 & 0 & 0 & 0 & 0 & 0
\\
 0 & 0 & 0 & 0 & -1 & 0 & 0 & 0
\\
 0 & 0 & 0 & 1 & 0 & 0 & 0 & 0
\\
 0 & 0 & 1 & 0 & 0 & 0 & 0 & 0
\\
 0 & 0 & 0 & 0 & 0 & 1 & 0 & 0
\\
 -1 & 0 & 0 & 0 & 0 & 0 & 0 & 0
\\
 0 & 0 & 0 & 0 & 0 & 0 & 0 & 1
\end{array}\right)$\\
\end{tabular}\\
\begin{tabular}{l@{\hspace{0.5cm}}l}
$(s_{4,6}|A_4) \izo (s_{4,6}|B2+A_1; \operatorname{P10}) $ & $
R=\left(\begin{array}{cccc}
0 & 1 & 0 & 0
\\
 -1 & 0 & 0 & 0
\\
 0 & 0 & 0 & 0
\\
 0 & 0 & 0 & 0
\end{array}\right)$\\
$(s_{4,11}|A_4) \izo  (s_{4,11}|B2+A_1; \operatorname{P10}) $ & $
R=\left(\begin{array}{cccc}
0 & \frac{1}{2} & 0 & 0
\\
 -\frac{1}{2} & 0 & 0 & 0
\\
 0 & 0 & 0 & 0
\\
 0 & 0 & 0 & 0
\end{array}\right)$\\
$(s_{4,11}|A_4) \izo (s_{4,11}|B2+A_1; \operatorname{P12})  $ & $
R=\left(\begin{array}{cccc}
0 & 0 & 1 & 0
\\
 0 & 0 & 0 & 0
\\
 -1 & 0 & 0 & 0
\\
 0 & 0 & 0 & 0
\end{array}\right)
$\\
$(s_{4,11}|A_4) \izo (s_{4,11}|s_{4,6}; \operatorname{P11})  $ & $
R=\left(\begin{array}{cccc}
0 & 0 & 0 & 0
\\
 0 & 0 & 0 & -1
\\
 0 & 0 & 0 & 0
\\
 0 & 1 & 0 & 0
\end{array}\right)$\\
$(s_{4,11}|A_4) \izo (s_{4,11}|s_{4,11}; \operatorname{P24}) $ & $
R=\left(\begin{array}{cccc}
0 & 0 & 0 & 1
\\
 0 & 0 & -1 & 0
\\
 0 & 1 & 0 & 0
\\
 -1 & 0 & 0 & 0
\end{array}\right)
$\\
\end{tabular}

\subsection*{Drinfeld double \arabic{DDCounter}}\addtocounter{DDCounter}{1}
\begin{tabular}{|c|c|c|c|l|}
\hline
Derived  & Lower central & Algebra of  &  Killing form & Manin triples \\
series  & series & derivations & signature  &\\
\hline
 \multirow{3}{*}{$\{8,7,1,0\}$} & \multirow{3}{*}{$\{8,7\}$} & \multirow{3}{*}{17} & \multirow{3}{*}{$\{1,7,0\}$}
 & {$(s_{4,3}^{1,1}|A_4)$}\\
 & & & & {$(s_{4,3}^{1,1}|B2+A_1; \operatorname{P10})$}\\
 & & & & {$(s_{4,3}^{1,1}|s_{4,3}^{1,-1}; \operatorname{P2})$}\\
\hline
\end{tabular}\\[8pt]
\begin{tabular}{l@{\hspace{0.5cm}}l}
$(s_{4,3}^{1,1}|A_4) \izo (s_{4,3}^{1,1}|B2+A_1; \operatorname{P10})$ & $ R=
\left(\begin{array}{cccc}
0 & \frac{1}{2} & 0 & 0
\\
 -\frac{1}{2} & 0 & 0 & 0
\\
 0 & 0 & 0 & 0
\\
 0 & 0 & 0 & 0
\end{array}\right)$\\
$(s_{4,3}^{1,1}|A_4) \izo (s_{4,3}^{1,1}|s_{4,3}^{1,-1}; \operatorname{P2})$ & $ R=
\left(\begin{array}{cccc}
0 & 0 & 0 & 0
\\
 0 & 0 & 0 & 0
\\
 0 & 0 & 0 & -1
\\
 0 & 0 & 1 & 0
\end{array}\right)$
\end{tabular}

\subsection*{Drinfeld double \arabic{DDCounter}}\addtocounter{DDCounter}{1}

\begin{tabular}{|c|c|c|c|l|}
\hline
Derived  & Lower central & Algebra of  &  Killing form & Manin triples \\
series  & series & derivations & signature  &\\
\hline
 \multirow{3}{*}{$\{8,7,1,0\}$} & \multirow{3}{*}{$\{8,7\}$} & \multirow{3}{*}{13} &
 $\{1,7,0\}$ for $a>\sqrt{2}$ & $(s_{4,5}^{a,0}|A_4)$\\
 & & & $\{0,8,0\}$ for $a=\sqrt{2}$ & $(s_{4,5}^{a,0}|B2+A_1; \operatorname{P10})$\\
 & & & $\{0,7,1\}$ for $a<\sqrt{2}$ & $(s_{4,5}^{a,0}|s_{4,5}^{a,0}; \operatorname{P22}))$\\
\hline
\end{tabular}\\[8pt]
\begin{tabular}{l@{\hspace{0.5cm}}l}
$(s_{4,5}^{a,0}|A_4) \izo (s_{4,5}^{a,0}|B2+A_1; \operatorname{P10})$ & $R=
\left(\begin{array}{cccc}
0 & \frac{a}{a^{2}+1} & \frac{1}{a^{2}+1} & 0
\\
 \frac{-a}{a^{2}+1} & 0 & 0 & 0
\\
 \frac{-1}{a^{2}+1} & 0 & 0 & 0
\\
 0 & 0 & 0 & 0
\end{array}\right)$\\
$(s_{4,5}^{a,0}|A_4) \izo (s_{4,5}^{a,0}|s_{4,5}^{a,0}; \operatorname{P22})$ & $R=
\left(\begin{array}{cccc}
0 & 0 & 0 & 1
\\
 0 & 0 & 0 & 0
\\
 0 & 0 & 0 & 0
\\
 -1 & 0 & 0 & 0
\end{array}\right)$
\end{tabular}

\subsection*{Drinfeld double \arabic{DDCounter}}\addtocounter{DDCounter}{1}
\begin{tabular}{|c|c|c|c|l|}
\hline
Derived  & Lower central & Algebra of  &  Killing form & Manin triples \\
series  & series & derivations & signature  & \\
\hline
 \multirow{4}{*}{$\{8,7,1,0\}$} & \multirow{4}{*}{$\{8,7\}$} & \multirow{4}{*}{11} & \multirow{4}{*}{\begin{tabular}{l}
$\{1,7,0\}$ for $a>\sqrt{\frac{2}{3}}$\\
$\{0,8,0\}$ for $a=\sqrt{\frac{2}{3}}$\\
$\{0,7,1\}$ for $a<\sqrt{\frac{2}{3}}$
\end{tabular}
}
 & $(s_{4,5}^{a,a}|A_4)$\\
 & & & & $(s_{4,5}^{a,a}|B2+A_1; \operatorname{P10})$\\
 & & & & $(s_{4,5}^{a,a}|B2+A_1; \operatorname{P22},\epsilon)$\\
 & & & & $(s_{4,5}^{a,a}|s_{4,5}^{a,-a}; \operatorname{P22})$\\
 \hline
\end{tabular}\\[8pt]
\begin{tabular}{l@{\hspace{0.5cm}}l}
$(s_{4,5}^{a,a}|A_4) \izo (s_{4,5}^{a,a}|B2+A_1; \operatorname{P10})$ & $R=
\left(\begin{array}{cccc}
0 & \frac{2 a}{4 a^{2}+1} & \frac{1}{4 a^{2}+1} & 0
\\
 \frac{-2 a}{4 a^{2}+1} & 0 & 0 & 0
\\
 \frac{-1}{4 a^{2}+1} & 0 & 0 & 0
\\
 0 & 0 & 0 & 0
\end{array}\right)$\\
$(s_{4,5}^{a,a}|A_4) \izo (s_{4,5}^{a,a}|B2+A_1; \operatorname{P22},\epsilon)$ & $R=
\left(\begin{array}{cccc}
0 & 0 & 0 & 0
\\
 0 & 0 & \frac{\epsilon}{2 a} & 0
\\
 0 & -\frac{\epsilon}{2 a} & 0 & 0
\\
 0 & 0 & 0 & 0
\end{array}\right)$\\
$(s_{4,5}^{a,a}|A_4) \izo (s_{4,5}^{a,a}|s_{4,5}^{a,-a}; \operatorname{P22})$ & $R=
\left(\begin{array}{cccc}
0 & 0 & 0 & 1
\\
 0 & 0 & 0 & 0
\\
 0 & 0 & 0 & 0
\\
 -1 & 0 & 0 & 0
\end{array}\right)$
\end{tabular}\\
Note that Drinfeld doubles 8 and 9 form one-parameter families (parametrized by $a$). So far, each family has been shown to contain at least three members, distinguished by the signatures of the Killing form.

%%%%%%%%%%%%%%%%%%%%%%%%%%%%%%%%%%%%%%%%%%%%%%%%%%%%%%%%%%%%%%%%%%%%%%%%%%%%%%%
%% Backmatter
%%%%%%%%%%%%%%%%%%%%%%%%%%%%%%%%%%%%%%%%%%%%%%%%%%%%%%%%%%%%%%%%%%%%%%%%%%%%%%%

\end{document}